\documentclass[12pt]{article}
\pdfoutput=1
\usepackage{putex}
\usepackage{graphicx}
\usepackage{epstopdf}
\usepackage{enumerate}
\usepackage[nosort]{cite}
\usepackage{tensor}
\usepackage{slashed}
\usepackage{feynmf}
\usepackage{hyperref}
\usepackage{float}
\usepackage[bottom]{footmisc}
\usepackage[utf8]{inputenc}
\usepackage{amsmath}
\usepackage[usenames,dvipsnames,svgnames,table]{xcolor}
\usepackage[bf,hang,footnotesize]{caption}

\numberwithin{equation}{section}
\hypersetup{
	pdfstartview={FitH},    % fits the width of the page to the window
	pdftitle={Title},    % title
	pdfauthor={Kim, Kraus, Myers},     % authors
	colorlinks=true,       % false: boxed links; true: colored links
	linkcolor=blue,          % color of internal links
	citecolor=blue!59!black! ,        % color of links to bibliography
	filecolor=magenta,      % color of file links
	urlcolor=blue!75!black,           % color of external links
	linktoc=all
}

\makeatletter
\g@addto@macro\bfseries{\boldmath}
\makeatother

\numberwithin{equation}{section}

\newcommand{\eq}[2]{\begin{align}\label{#1}#2\end{align}}

\def\wb{\overline{w}}
\def\Lc{{\cal L}}

\newcommand {\be} {\begin {equation}}
\newcommand {\ee} {\end {equation}}

\newcommand{\p}{\partial}

\def\Tr{\mathop{\rm Tr}}
\def\tr{\mathop{\rm tr}}
\newcommand\rt{{\rightarrow}}
\def\eps{\epsilon}

\newcommand{\rf}[1]{(\ref{#1})}
\newcommand{\rff}[1]{\ref{#1}}

\newcommand{\Fb}{\overline{F}}

\newcommand{\Lb}{\overline{L}}

\definecolor{greenC}{rgb}{0.0, 0.38, 0.18}

\newcommand{\RR}{\mathbb{R}}

\newcommand{\Ab}{\overline{A}}

\newcommand{\ab}{\overline{a}}

\newcommand{\dr}{\text{d}}
\newcommand{\eL}{\mathcal{L}}
\newcommand{\Kh}{\hat{K}}
\newcommand{\Lch}{\hat{\Lc}}
\newcommand{\kb}{\overline{k}}

\begin{document}

\institution{UCLA}{ \quad\quad\quad\quad\quad\quad\quad\ ~ \, $^{1}$Mani L. Bhaumik Institute for Theoretical Physics
		\cr Department of Physics \& Astronomy,\,University of California,\,Los Angeles,\,CA\,90095,\,USA}

\title{ Systematics of Boundary Actions in Gauge Theory and Gravity
	}

\authors{Seolhwa Kim$^{1}$, Per Kraus$^{1}$, Richard M. Myers$^{1}$}
	
\abstract{We undertake a general study of the boundary (or edge) modes that arise in gauge and gravitational theories defined on a space with boundary, either asymptotic or at finite distance, focusing on efficient techniques for computing the corresponding boundary action.   Such actions capture all the dynamics of the system that are implied by its asymptotic symmetry group, such as correlation functions of the corresponding conserved currents.  Working in the covariant phase space formalism, we develop a collection of approaches for isolating the boundary modes and their dynamics,   and illustrate with various examples, notably AdS$_3$ gravity (with and without a gravitational Chern-Simons terms) subject to assorted boundary conditions. 
 }
	
	\date{}
	
	\maketitle
	\setcounter{tocdepth}{2}
	\begingroup
	\hypersetup{linkcolor=black}
	\tableofcontents
	\endgroup
	
%%%%%%%%%%%%%%%%%%%%%%%%%%%%%%%%%%%%%%%%%%%%%%%%%%%

\section{Introduction}

The purpose of this paper is to systematically understand the mechanism by which gauge theories\footnote{We use the term ``gauge symmetry" to denote any local symmetry, including Yang-Mills and diffeomorphisms.} defined on spaces with boundaries (either at finite distance or asymptotic) are found to host local degrees confined to the boundary.   These boundary degrees of freedom are governed by a boundary action and we aim to develop general and efficient methods for calculating it.  The simplest and most familiar example  is provided by the Chern-Simons/WZW correspondence \cite{Witten:1988hf,Elitzur:1989nr}.  

The existence of boundary modes is tied to the crucial distinction between those gauge symmetries that act nontrivially at the boundary versus those which are suitably localized away from the boundary.   The former take the system from one point in phase space (or from one quantum state) to another, while the latter do not.  We refer to these as large and small gauge transformations respectively.   For  example, in the case of General Relativity in asymptotically flat spacetime two black hole configurations, one at rest and one in uniform motion, are related by a large coordinate transformation.   Large gauge transformations are generated by nontrivial conserved charges, and the large versus small distinction implies that such charges can be expressed as  boundary integrals.  This fact has as its most elementary incarnation the  flux integral expression for charge in electrodynamics, and finds its most general  expression in the covariant phase space formalism \cite{Wald:1999wa}.  

Of interest to us in this work are cases in which the large gauge transformations are associated to physical degrees of freedom localized at the boundary.  In a canonical formulation, physical degrees of freedom are nonzero modes of the symplectic form.  In a gauge theory the symplectic form breaks up into bulk and boundary pieces.  Boundary degrees of freedom are zero modes of the bulk part of the symplectic form, but not of the boundary part.  The boundary modes are furthermore governed by a boundary action.  Particularly interesting is the case in which the group of large gauge transformations, i.e the asymptotic symmetry group, is infinite dimensional, in which case there is a boundary field theory.  One motivation for our work is that  these boundary field theory degrees of freedom are every bit as physical as any other and so should be understood.   Although we do not discuss it here, boundary modes also play an import role in attempts to formulate entanglement entropy in gauge theory and gravity; a few  references include \cite{Donnelly:2014fua,Ghosh:2015iwa,Donnelly:2016auv,Blommaert:2018oue}.   Discussions of boundary modes in various other contexts include   \cite{Compere:2015knw,Seraj:2017rzw,Kutluk:2019ghr,Adami:2020ugu,Adami:2021sko}.

In the case of  Chern-Simons (CS) theory there is a simple well known procedure for obtaining the WZW boundary action \cite{Elitzur:1989nr}.  In CS theory the Gauss law constraint is the statement that the spatial components of the field strength vanish.  The general solution to this constraint is obtained by writing the gauge fields as a gauge transformation of a given flat connection.  Inserting this back into the action the CS Lagrangian becomes a total derivative and the resulting boundary term is the WZW action.\footnote{To get a fully explicit two-dimensional boundary action one needs to choose an explicit parametrization of the gauge group elements, as we review in Appendix \ref{Non-Abelian CS}.}   

This strategy is not necessarily so easy to carry out in other theories, in particular if there is no easy way to find the general solution of the constraints.  A main objective of this paper is to develop more widely applicable methods for deducing boundary actions, illustrated by explicit examples.    To this end we primarily work in a  covariant phase space framework \cite{Crnkovic:1986ex,Lee:1990nz,Wald:1999wa},  where the main actors are the symplectic form and the boundary charges.   We take as examples three-dimensional gravity, possibly supplemented with a gravitational CS term, subject to various boundary conditions.  As is well known \cite{Achucarro:1986uwr,Witten:1988hc}, pure 3D  gravity with AdS$_3$ boundary conditions admits a CS formulation to which the basic CS/WZW procedure above can be applied (with some modifications due to the change of boundary conditions), and one obtains a theory of boundary gravitons \cite{Coussaert:1995zp,Cotler:2018zff,Barnich:2017jgw}.    The metric formulation version of this procedure was worked out in \cite{Kraus:2021cwf,Ebert:2022cle}, and extended to the case of a finite cutoff boundary.  Here we illustrate our more general methods by considering non-AdS boundary conditions, including those of warped AdS$_3$ \cite{Anninos:2008fx} supported by the gravitational CS term \cite{Deser:1981wh}; note that in general the addition of the gravitational CS term breaks the correspondence with the usual SL$(2,\RR)\times {\rm SL}(2,\RR)$ CS formulation.  

We now briefly summarize our various approaches to deriving boundary actions.  In the canonical approach followed here, the boundary action is built out of a canonical 1-form $\Theta$, whose exterior variation yields the symplectic form $\Omega$, and the boundary Hamiltonian. These objects are defined on a phase space which consists of a particular gauge orbit, obtained by acting on a chosen background solution with  all possible  boundary  condition preserving gauge transformations.   The basic point is that on a given orbit the symplectic form as well as the charges generating the large gauge transformations (i.e.  asymptotic symmetries) all localize to the boundary.   If the boundary charges $Q_V$ associated to large gauge transformations $V$ can be computed, the symplectic form may in principle be found by solving the equation $i_V\Omega = - \delta Q_V$.  Solving this can be quite laborious, and one still needs to compute the potential $\Theta$ for $\Omega$.  For the class of examples that arise in  AdS$_3$ gravity with various boundary conditions, it is possible to bypass all this and pass directly from the conserved (angular)momentum and Hamiltonian to the boundary action, as was employed in the example of cutoff AdS$_3$ gravity in \cite{Ebert:2022cle} leading to a boundary Nambu-Goto action, whose origin was  clarified in \cite{Kraus:2022mnu}.   

 As this approach may not always be possible, we also develop more general methods for computing $\Omega$.   These are  based on identifying a phase space  1-form valued vector field $W$, which we refer to as the transfer field, which obeys the relation $\delta \phi = i_W \delta \phi$ when $\delta \phi$ is restricted to a single gauge orbit.  Given knowledge of $W$ and of the charges $Q$ we show how one can use these to read off the boundary symplectic form. Furthermore, we demonstrate another technique which is somewhat less efficient for computing $\Omega$, but has the advantage of allowing one to sometimes obtain expressions for the boundary contributions to $\Omega$ independent of the chosen boundary conditions. We describe when this can be done, most notably for diffeomorphisms. As an example, the Einstein-Hilbert action in any dimension with any cosmological constant always produces the contribution \eqref{19 examples} to the boundary symplectic form, independent of the boundary conditions.

 The main example we use to illustrate our general methods is warped AdS$_3$ \cite{Nutku:1993eb,Gurses:1994bjn,Anninos:2008fx}.  This is a well-studied  solution of topologically massive gravity (TMG) \cite{Deser:1981wh}.  The warped asymptotics make it less obvious a priori on what  surface the boundary action should be thought of as living.   We will work out the boundary action in detail, and also show how the same results may be obtained via ``lower spin gravity" \cite{Hofman:2014loa} which is a CS formulation that can be used to describe a  subsector of the full TMG phase space.     

These boundary actions are important inasmuch as the the boundary modes are part of the dynamical degrees of freedom of the theory.  For example the boundary photons and gravitons arising in the CS and AdS$_3$ gravity theories contribute to the thermal partition function \cite{Maloney:2007ud}.  However, at first sight the physical relevance of these modes may seem elusive, given that they are generated by performing gauge transformations.  This point is clarified by coupling another system to the theory containing the boundary modes.  We give a simple example of this in which a boundary scalar couples to a CS theory defined on a spatial disk, showing how correlators of the boundary scalar are modified by the coupling to the boundary photons.

Another situation occurs when the boundary is not the true ``end" of the spacetime, but rather an interface  marking a transition  between two different regions with distinct asymptotics.    It is interesting to ask whether and how the  modes that ``would have been there" had the interface been an actual boundary manifest themselves in the full system.   One can think of this as a version of the setup described in the previous paragraph, where one side of the transition region now functions as the additional system.   This situation arises very naturally in gravity.  For example, one can have a solution with a near horizon AdS$_3$ region, which by itself supports boundary modes, embedded inside an asymptotically AdS$_{D>3}$ solution.  The latter solution has a finite dimensional asymptotic symmetry group, so one may wonder whether the near horizon boundary modes are detectable at the asymptotic boundary.  We answer this question in the affirmative, showing how the pure gauge modes in the near horizon are promoted to non-pure gauge modes in the full spacetime.

It is worthwhile to clarify our usage of certain terminology in what follows.  In particular, when we refer to a gauge transformation, depending on context we may or may not distinguish whether we mean small or large gauge transformations, and likewise for diffeomorphisms.  When this matters, which is often, we will distinguish the two. Recall that small gauge transformations/diffeomorphisms describe redundancies, and in a canonical framework are zero modes of the symplectic form.  Large gauge transformations/diffeomorphisms instead move us between distinct points in phase space, and are nonzero modes of the symplectic form.  Finally, we occasionally use the term ``local symmetry", which is meant to encompasses both gauge and diffeomorphism symmetry, whether small or large. The usage should always be clear from the context.

The rest of this paper is organized as follows.  In Section \rff{bndyacts}, after quickly reviewing the standard approach to U(1) CS theory on a disk and the physical relevance of boundary modes, we go on to a general discussion of the origin and identification of boundary modes within the framework of the covariant phase space.  In Section \rff{compute} we develop specific methods for computing boundary actions, illustrated through particular examples. In Section \rff{Application to Topologically Massive Gravity} we consider the case of warped AdS$_3$ asymptotics in topologically massive gravity, which is a useful and nontrivial example to illustrate various issues.  We obtain the boundary action in both the metric formulation and in the so-called lower spin gravity formulations.  Section \rff{emergent} discusses how boundary modes can appear in the IR.  A series of appendices lay out some conventions, review an important theorem regarding identically closed forms, review the proper method for handling  non-diffeomorphism invariant actions  in the Wald formalism, review non-Abelian CS theory and apply our methods to it, and explain the connection of our 3D gravity results to 2D JT gravity.

\section{Review of Boundary Actions}
\label{bndyacts}

In this section we discuss  general aspects of boundary modes, their origin in terms of large gauge transformations, and  the construction of an action that describes them.

\subsection{\texorpdfstring{$U(1)$}{U(1)} CS theory on \texorpdfstring{$M=D \times \RR$}{M = DxR}}

To get oriented, we first  quickly review the simple and classic example of Chern-Simons theory on a spatial disk and the corresponding  boundary gauge modes, following the original Lagrangian approach \cite{Elitzur:1989nr}.  This approach is based on solving the Gauss law  constraint and substituting back into the action, yielding a total derivative.  While this method works well here, it is not so easy to adapt to other examples such as gravity in the metric description.  For this reason we go on to develop  more flexible methods based on a covariant phase space analysis. 

The action for for Abelian CS theory on a spatial disk cross time is
\eq{aa}{S & = k \int_M A\wedge  \dr A + S_{bndy}\cr
& =  -k \int_M\! \dr^3x (A_r \p_t A_\phi - A_\phi \p_t A_r +2A_t F_{\phi r} ) +S'_{bndy}~,}
where we integrated by parts and absorbed the boundary term into $S'_{bndy}$.  We choose boundary conditions $\delta (A_t-A_\phi)|_{\p M}=0$. 

Specializing to the case $(A_t-A_\phi)|_{\p M}=0$,  a good variational principle is achieved by taking
\eq{ab}{S_{bndy} = 0,\ \ \ \ S'_{bdy}  =-k \int_{\p M} \dr t\dr \phi A_\phi^2~.}
$A_t$ is  a Lagrange multiplier enforcing the constraint $F_{\phi r}=0$, which is solved by writing 
\eq{ac}{ A_\phi  = \p_\phi \alpha~,\quad A_r = \p_r \alpha~,}
with $\alpha(r,t,\phi+2\pi)= \alpha(r,t,\phi)$.  Plugging this back into the action, the bulk terms become a total derivative, and we arrive at the chiral boson action
\eq{ad}{S = k \int_{\p M} \dr t \dr\phi (\p_\phi \alpha \p_t \alpha - \p_\phi \alpha \p_\phi \alpha)~.}
The basic equal time  Poisson-Dirac bracket is 
\eq{af}{ \{ \alpha(\phi),\p_\phi \alpha(\phi')\}= \frac{1}{ 2k} \delta (\phi-\phi')~.}
The Hamiltonian and charges generating infinitesimal gauge transformations by $\lambda$  are 
\eq{ae}{ H_t = k \int_0^{2\pi} (\p_\phi \alpha)^2 \dr\phi,\ \ \ \ H[\lambda] = 2k\int_0^{2\pi}\lambda \p_\phi \alpha  \dr\phi~.}
The theory describes  a $U(1)$ current $J=k\p_\phi \alpha $ whose Fourier modes obey a $U(1) $  current algebra.  The current-current-correlator is 
\eq{afa}{ G_{JJ}(w)= \langle J(w) J(0)\rangle = -\frac{k}{ 4 \pi\sin^2\left(\frac{w}{2}\right)}~,\quad  w =\phi + t, \quad \overline w = \phi - t~.  }

\subsubsection{Physical relevance of boundary modes}

Inasmuch as the preceding analysis shows that boundary modes can carry nonzero energy and momentum, they are established as being nontrivial physical states.   Nonetheless, their ``pure gauge" character leads one to wonder, at least upon first hearing, whether they might be ignorable in some sense, for example by decoupling from the rest of the physical system in which they are embedded.  However, it is not hard to show that the boundary modes do have measurable consequences on other observables.

To expose these effects  we can think of coupling the theory in the  disk region to some external system comprised of charged matter that couples to the CS gauge field at the boundary of the disk.  In the simplest incarnation we can take the system to live on the boundary of the disk, and to be completely explicit we consider a charged scalar field example,
\eq{tai}{S = k\int_M A\wedge \dr A  + \int_{\p M} \dr ^2x  (D^\mu \Phi)^* D_\mu \Phi ~.}
The covariant derivative is taken to correspond to a gauging of the scalar shift symmetry, $D_\mu \Phi = \p_\mu \Phi -iq A_\mu$.    This is convenient, since the associated current, $J^\Phi_\mu = i(\p_\mu \Phi - \p_\mu \Phi^*)$  decomposes into dimension $(1,0)$ and $(0,1)$ operators, which is not the case for the current associated to phase rotations of the scalar.\footnote{The non-invariance of this current  under large gauge transformations is not a concern  since such transformations are to be viewed as global symmetries.} Repeating the previous steps we arrive at the action
\eq{tao}{ S =  \int \dr^2x\left[ -2k \p_\phi \alpha \p_{\overline w} \alpha  +4 \p_w \Phi^* \p_{\overline w} \Phi +2iq  \p_\phi\alpha (  \p_{\overline w} \Phi - \p_{\overline w} \Phi^*) \right]  }
with  $\p_w = \frac{1}{2}(\p_\phi + \p_t), \p_{\overline w} = \frac{1}{2}(\p_\phi - \p_t)$. The coupling of $\alpha$ to $\Phi$ has no effect on the energy spectrum of the theory, as follows from the fact that the coupling can be removed by a redefinition of $\Phi$. There is a nontrivial effect on scalar correlators, in particular on the two-point function of the  current $J^\Phi_{\overline w}$.   This effect reflects the fluctuating phase acquired by a charged particle on traveling between the two operator insertion points on the boundary.   Treating $q$ as a perturbation, we can readily sum up contributing diagrams by performing Wick contractions, resulting in
\eq{tap}{ G^{(q)}_{J^\Phi J^\Phi}(p) &=  \langle J^\Phi_{\overline{w}}(p) J^\Phi_{\overline{w}}(-p)\rangle_q \cr
& = \sum_{m=0}^\infty  (-4q^2)^m  [G_{J^\Phi J^\Phi}(p)]^{m+1}  [ G_{\alpha\alpha}(p)]^m \cr
& =  \frac{G_{J^\Phi J^\Phi}(p)}{1+ 4q^2 G_{J^\Phi J^\Phi}(p) G_{\alpha \alpha}(p)}~,}
where the $q=0$ correlators are
\eq{tapa}{ & G_{J^\Phi J^\Phi}(p)  = \langle J^\Phi_{\overline w}(p)J^\Phi_{\overline w}(-p)\rangle \sim \frac{p_{\overline w} }{p_w } \cr
& G_{\alpha \alpha }(p)  = \langle \p_\phi \alpha (p)\p_\phi \alpha (-p)\rangle \sim \frac{p_\phi }{ k p_{\overline w}} ~. }
At $q=0$ the correlator behaves as $p_{\overline w}/p_w$ corresponding to $1/\sin^2(\frac{\overline w}{2})$ in position space.  As $q\rt \infty$ the correlator tends to zero, with leading behavior $\frac{k}{q^2} \frac{p_{\overline w}}{p_\phi}$; being polynomial in $p_t$, this vanishes for unequal times.  

 The point to be emphasized here is that the boundary modes leave a  detectable imprint on the scalar correlators, and so are clearly ``real."

\subsection{Review of covariant phase space formalism} \label{Sec: cov ps formalism}

We begin with a brief review of the covariant phase space formalism, which will also serve to establish notation for the remainder of this paper. Along the way, we  make  comments about precisely where boundary conditions enter the formalism, as these will be useful to keep in mind later. For a more detailed review, see for example \cite{Crnkovic:1986ex,Wald:1999wa,Compere:2018aar,Harlow:2019yfa}.

\subsubsection{Action and covariant phase space}

We consider a theory defined on a $D = d + 1$ dimensional spacetime $M$ which admits a foliation by codimension-1 slices which we will generally denote by $\Sigma$. The, potentially asymptotic, boundary structure of $M$ can then be decomposed into $\p M = \Sigma_+ \cup \Gamma \cup \Sigma_-$ where $\Sigma_\pm$ are the slices in the asymptotically far future and past and $\Gamma$ is formed by unioning the boundaries of all the slices. On this spacetime we consider a theory with fields $\phi$ whose dynamics are described by an action
\eq{1 canonical}{
S[\phi] = \int_M L + \int_{\p M}\ell
}
where $L$ is the $D$-form Lagrangian and $\ell$ is some allowed $d$-form boundary contribution\footnote{Though we use the same symbol $\ell$ for the boundary contribution over all of $\p M$, there need be no relation between $\ell$ on $\Gamma$ and $\ell$ on $\Sigma_\pm$. Shifts in this $\ell$ on either always produce shifts in the canonical 1-form by something $\delta$ exact and so do not change the symplectic structure.}, which we assume to be local functionals of the fields.

Throughout we will use $\delta$ to denote the exterior variational derivative, which we refer to as the variation. Particular infinitesimal transformations of the fields will be thought of as vector fields $V$ on field space. The action of a vector $V$ on a field $\phi$ will then be denoted by the contraction\footnote{Though here we prefer the contraction on variation notation to denote infinitesimal transformation, the reader may find it helpful to recall that the following are equivalent: $i_V\delta F[\phi] = \eL_VF[\phi] = V(F[\phi])$ where $\eL_V$ denotes the Lie derivative on field space and $V(F[\phi])$ is the action of the vector field $V$ on the function $F$ on field space.} $i_V\delta \phi$. By integrating by parts, the variation of the Lagrangian may always be written
\eq{2 canonical}{
\delta L = E \wedge \delta \phi + \dr \theta
}
for some $\theta$. Setting $E = 0$ will be our equations of motion. We note that $\theta$ is always ambiguous up to addition of a $\dr$-closed form which we will return to shortly\footnote{In the literature, e.g. \cite{Wald:1999wa}, it is often mentioned that $\theta$ is also ambiguous up to addition of a $\delta$-closed form. While this is true, any such shift in $\theta$ is equivalent to shifting $\ell$. So we take the perspective that $\theta$ has no $\delta$ ambiguity, but $\ell$ remains to be chosen.}.

Using this identity, the variation of the action is given by
\eq{3 canonical}{
\delta S = \int_M E\wedge \delta \phi + \int_{\Sigma_+-\Sigma_-}(\theta + \delta\ell) + \int_\Gamma(\theta + \delta\ell).
}
In order to have a good variational principle we require that the on-shell variation have no support on $\Gamma$ which then requires
\eq{4 canonical}{
(\theta + \delta\ell)|_\Gamma = \dr B
}
for some $B$. Of course, this $B$ can always be absorbed into a redefinition of $\theta$. However, we note that for a generic theory, the LHS above will not automatically take the form of a total derivative. Instead, there will be terms which only vanish upon the imposition of boundary conditions. This means that the $B$ here generally depends on the boundary conditions we choose for our theory, and the existence of $B$ may impose conditions on what we choose for $\ell$\footnote{A standard example of this would be the need to include the Gibbons-Hawking-York term in the Einstein-Hilbert action with Dirichlet boundary conditions, though in that case we may choose $B = 0$ depending on our gauge fixing, see e.g. \cite{Harlow:2019yfa} for details.}. As a simple example, starting from the action \eqref{aa} we find
\eq{afa-1}{
\theta = -k A \wedge \delta A, \ \ \ \ B = 0
}
with chiral boundary conditions.

It will be useful later  to keep explicit which objects depend on the boundary conditions and which do not. So while it is possible to absorb $B$ into a redefinition of $\theta$, we will refrain from doing so in order to avoid reference to boundary conditions when writing $\theta$.

It's also useful to observe that having a good variational principle is equivalent to slice independence of the symplectic form. The potential for the symplectic form is always found by extracting what remains of the action's variation from the initial and final time slices; we write $\mathcal{A}$ to denote this.  Here we have
\eq{4-1 canonical}{
\delta S = \int_{\Sigma_+ - \Sigma_-}\mathcal{A} = \int_{\Sigma_+ - \Sigma_-}(\theta + \delta\ell - \dr B)
}
after the imposition of \eqref{4 canonical}. This means we should choose our symplectic form to be\footnote{Throughout this work we ignore the complications that come from the possibility of non-trivial phase space topology, including the possibility of a symplectic form with non-trivial De Rahm cohomology.}
\eq{4-2 canonical}{
\Omega = \int_\Sigma \omega = \int_\Sigma \delta\mathcal{A} = \int_\Sigma \delta(\theta - \dr B).
}
Though the ultimate argument for using this object as our symplectic form will be that it produces the desired Poisson brackets, we can see an immediate benefit by taking a second variation of \eqref{4-1 canonical}, which implies that this $\Omega$ is independent of the slice we choose.

It is, however, useful to observe that we can show more directly that the slice independence of $\Omega$ is precisely equivalent to the demand \eqref{4 canonical}, and hence the demand for a good variational principle. To see this we take a second variation of \eqref{2 canonical} to find
\eq{6 canonical}{
-\delta(E\wedge \delta\phi) = \dr\delta\theta = \dr\omega
}
so the symplectic current is closed on-shell. Hence $\Omega(\Sigma)$ is independent of the slice $\Sigma$ if and only if the pullback of $\omega$ to $\Gamma$ vanishes. This demand can be rewritten as $\omega|_\Gamma = \delta(\theta|_\Gamma - \dr B) = \delta[(\theta + \delta \ell)|_\Gamma - \dr B]$ so the condition \eqref{4 canonical}, obtained from demanding a good variational principle, is equivalent\footnote{Strictly speaking, the slice independence of the symplectic form only implies \eqref{4 canonical} up to a $\delta$-closed form, but locally on phase space such a form can be written as exact and absorbed into a redefinition of $\ell$.} to the slice independence of the symplectic form.

As a final comment about this definition for the symplectic form, we should discuss the distinction between the phase space and prephase space. We will generally take prephase space to consist of all configurations of the fields which obey the boundary conditions and the equations of motion. On this space the quantity \eqref{4-2 canonical} will generally be degenerate and hence cannot be a proper symplectic form\footnote{This is closely related to the lack of deterministic evolution on prephase space; specifying the fields and some number of their derivatives on a Cauchy slice may not uniquely determine the same data on a later time slice. The most common way to exhibit this non-uniqueness is by specifying a configuration and applying to it a gauge transformation which differs from the identity only at times later than the first slice. We would thus have two solutions to the equations of motion whose data agree on one slice but disagree on another.}. For this reason, it's often referred to as the presymplectic form.

To form the actual phase space, we need to perform a symplectic quotient and mod out the null directions of the symplectic form. This is a matter of viewing the prephase space as a bundle whose fibers are the null directions and whose base space is our true phase space. The mathematical details were laid out in \cite{Lee:1990nz}, but in practice the result is that $\Omega$ is a non-degenerate 2-form on the base space and so working on the true phase space is a matter is ignoring those variables whose variation lies along the pure gauge directions\footnote{If we are being strict, this is the statement that, at least locally, a section of the bundle is diffeomorphic to the base space.}.

\subsubsection{Symmetries and charges}\label{Symmetries and Charges}

With the covariant phase space framework now in place, it will be important for us to review how symmetries enter the picture. A vector field $V$ on phase space is typically defined to be a symmetry if its action on the Lagrangian is a total derivative:
\eq{7 canonical}{
i_V\delta L = \dr k_V
}
for some $k_V$. Contracting $V$ onto \eqref{2 canonical} it now follows that
\eq{7-1 canonical}{
\dr J_V \equiv \dr(i_V\theta - k_V) = E\wedge i_V\delta\phi.
}
So $J_V = i_V\theta - k_V$ is the conserved Noether current associated to $V$.

Though any vector field $V$ satisfying \eqref{7 canonical} admits a conserved Noether current, constructing the Noether charge is not always as a simple as integrating the current over a time slice. There may be non-trivial boundary contributions to the true Noether charge $H_V$ in order for it to satisfy
\eq{10 canonical}{
i_V\Omega = -\delta H_V.
}

For general symmetries, one must directly evaluate the contraction on the symplectic form, but for gauge symmetries we may find the charges by another, sometimes more efficient, method. This was pointed out in \cite{Wald:1993nt} for the special case of diffeomorphism charges in diffeomorphism invariant theories, but with the Theorem \ref{Wald Theorem} of Appendix \rff{ident} it's simple to generalize this calculation to any gauge transformation, as we review now\footnote{Some additional simplifications that can help in computations are possible in the special case of diffeomorphisms even when the theory is not diffeomorphism invariant, as when gravitational Chern-Simons terms are included in the action. This still makes use of Theorem \ref{Wald Theorem} and was pointed out in \cite{Tachikawa:2006sz}. We review it in Appendix \ref{Charge Calculation}.}.

We suppose $V_\lambda$ generates a gauge transformation with gauge parameter $\lambda$, defined such that $V_0 = 0$ so $\lambda = 0$ is the identity transformation. Taking an additional variation of \eqref{7-1 canonical} it follows that, on-shell,
\eq{11 canonical}{
\dr(\eL_{V_\lambda}\theta - \delta k_\lambda) = 0
}
where we have abbreviated $k_{V_\lambda} = k_\lambda$. Thus we have a form closed for all free functions $\lambda$ and Theorem \ref{Wald Theorem} tells us that there must exist a phase space 1-form $\Pi_\lambda$, constructed locally from the fields and $\lambda$, such that
\eq{12 canonical}{
\eL_{V_\lambda}\theta - \delta k_\lambda = \dr \Pi_\lambda.
}
With this, it now follows from \eqref{4-2 canonical} that
\eq{13 canonical}{
\delta J_\lambda = -i_{V_\lambda}\omega + \dr(\Pi_\lambda - i_{V_\lambda}\delta B).
}
Thus if there exists a function $C_\lambda$ such that
\eq{14 canonical}{
\Pi_\lambda - i_{V_\lambda}\delta B = \delta C_\lambda,
}
\eqref{13 canonical} implies
\eq{15 canonical}{
i_{V_\lambda}\Omega = -\delta\int_\Sigma(J_\lambda - \dr C_\lambda).
}
We note that the existence of $C_\lambda$ is not guaranteed and will typically depend on the boundary conditions chosen for the theory\footnote{The insufficiency of \eqref{7 canonical} alone to ensure the existence of a charge satisfying \eqref{10 canonical} is pointed out many placed in the literature. In e.g. \cite{Wald:1999wa, Compere:2018aar}, integrability conditions are required as we see here and in \cite{Harlow:2019yfa} an auxiliary condition, there eq. (4.16), is required.}. The expression \eqref{15 canonical} now identifies the correct boundary Noether charge as being the integral of $J_\lambda$ with some additional boundary contributions.

Since $J_\lambda$ is closed for all $\lambda$, and is linear in $\lambda$, we can go further and compute a local functional, referred to as the Noether-Wald charge, $Q_\lambda$ such that $J_\lambda = \dr Q_\lambda$. With this the Noether charge may be written
\eq{16 canonical}{
H[\lambda] = \int_{\p \Sigma}(Q_\lambda - C_\lambda)
}
which has support only on the boundary of our Cauchy slice. Furthermore, we note that the only place the boundary conditions enter into this expression is through $C_\lambda$, as $J_\lambda$ and $Q_\lambda$ depend only on the Lagrangian of the theory.

Since the charges \eqref{16 canonical} have support only on the boundary, it follows immediately that any gauge transform whose parameters $\lambda$ have compact support away from any boundaries must produce vanishing Noether charge. The vector fields generating these transformations are thus identified from \eqref{15 canonical} as null directions of the presymplectic form which need to be modded out in the symplectic quotient. The non-zero Noether charges generate the large gauge transformations of the theory and are evidently localized to the boundaries of the spacetime. In the case these boundaries are asymptotic, the large gauge transformations are said to be asymptotic symmetries.

\subsection{Where boundary actions come from}\label{Where Boundary Actions Come From}

Using the machinery of the covariant phase space we can understand the, rather weak, sufficient conditions for producing boundary modes and gain some insight into the conditions under which the action for the theory is supported exclusively on the boundary. To transform this question into one which is easier to work with we first recall the phase space action. Writing the symplectic form for the theory as $\Omega = \delta\Theta$,  the phase space action is given by\footnote{Of course, $\Theta$ is not unique. Different $\Theta$ correspond to holding different data fixed in the initial and final configurations. For a point particle, $\Theta = p \delta x$ is the correct potential for varying with the initial and final position of the particle held fixed.}
\eq{1 where}{
S[\gamma] = \int_\gamma\Theta - \int_\gamma H_t\dr t
}
where $\gamma$ is some path through phase space parametrized by $t$. To orient ourselves it's useful to recall that for point particles $\Theta = p \delta x$ so
\eq{2 where}{
S = \int p\delta x - \int H_t \dr t = \int(p \dot x - H_t)\dr t
}
where the dot denotes the derivative along the path.

If time translation is a local symmetry, as it is in a diffeomorphism invariant theory, then by Theorem \ref{Wald Theorem} the Hamiltonian generating time translation $H_t$ is supported on the boundary. While this gives a boundary contribution to the action, we would still be left asking about when we receive boundary contributions from $\Theta$ and about what happens when time translation is not a local symmetry. We can obtain a better characterization of boundary contributions to the action which will help answer both of these questions by first introducing some notation.

Consider a generic theory with fields $\phi$ and some gauge group. We denote the action of a gauge group element with gauge parameters\footnote{For later convenience we assume that $\alpha = 0$ is the identity transformation and that these $\alpha$ are free functions on spacetime as might be obtained by exponentiating a Lie algebra about the identity.} $\alpha$ on $\phi$ by $T_\alpha[\phi]$. For example, if $\phi$ is a complex scalar field of charge $q$, $T_\alpha[\phi] = e^{iq\alpha}\phi$. If $\phi$ is a $U(1)$ connection we would have $T_\alpha[\phi] = \phi+ d\alpha$.

Now, we are always free to consider the field redefinition to gauge orbit variables where we pick a class of gauge-fixed configurations $\overline \phi$ so a general configuration is in the gauge orbit of some $\overline\phi$: $\phi = T_\alpha[\overline \phi]$. This separates the prephase space into gauge directions, parametrized by the $\alpha$, and non-gauge directions, parametrized by the gauge inequivalent $\overline\phi$.

In these variables if we write\footnote{In this section only we assume for convenience that $J_t$ has been defined to include all necessary boundary contributions already so $H_t$ is the full Noether charge satisfying \eqref{10 canonical}.} $H_t = \int_{\Sigma}J_t$, $J_t[\overline\phi, \alpha]$ is a local function of $\alpha$. Since $J_t$ is closed for all functions $\alpha$, Theorem \ref{Wald Theorem} tells us that we can construct a local potential $Q_t[\overline\phi, \alpha]$ such that
\eq{1-1 where}{
J_t[\overline\phi, \alpha] = J_t[\overline\phi, 0] + \dr Q_t[\overline\phi, \alpha].
}
That is to say the current separates into a component we would find if we had immediately fixed the gauge, and a boundary contribution which contains all the effects of the gauge modes $\alpha$, though it's notably also free to depend upon boundary excitations in the gauge fixed $\overline\phi$ directions. We note that $J_t[\overline\phi, 0] = 0$ when time translation is a gauge symmetry.

A similar argument works on the symplectic form. There are only three types of components that can appear in the symplectic form when we go to gauge orbit variables. This fixes the generic form of the symplectic form to be
\eq{1-2 where}{
\omega = \omega_{\overline\phi\overline\phi}[\overline\phi, \alpha] \delta \overline\phi \wedge\delta\overline\phi + \omega_{\overline\phi\alpha}[\overline\phi,\alpha]\delta\overline\phi\wedge\delta\alpha + \omega_{\alpha\alpha}[\overline\phi, \alpha] \delta\alpha\wedge\delta\alpha
}
where the coefficients should generally be understood to contract on any indices we have suppressed in $\delta\overline\phi$ and $\delta\alpha$, and may also contain derivatives that operate on the field variations.

Since the $\delta\alpha$ are completely free function in the bulk and $\dr\omega = 0$ on-shell by \eqref{6 canonical}, we may again apply theorem \ref{Wald Theorem} but now using $\delta\alpha$ instead of $\alpha$ as our free functions. We thus conclude\footnote{Less formally, one can understand that there ultimately cannot be any $\delta\alpha$ components appearing in the bulk or $\Omega$ would have non-zero contractions on small gauge transformations.}
\eq{1-3 where}{
\omega = \omega_{\overline\phi\overline\phi}[\overline\phi,\alpha]\delta\overline\phi \wedge\delta\overline\phi + \dr\tilde\omega_b[\overline\phi, \alpha]
}
where we have put a tilde on $\tilde\omega_b$ because it will not be the complete contribution to the boundary symplectic form. Importantly, $\tilde\omega_b$ is a local functional of the fields. 

We can go further by invoking theorem \ref{Wald Theorem}, this time on the $\alpha$ dependence of $\omega_{\overline\phi\overline\phi}[\overline\phi, \alpha]\delta\overline\phi\wedge\delta\overline\phi$. The result is evaluating $\omega_{\overline\phi\overline\phi}$ at $\alpha = 0$ and an additional, locally constructed contribution to the boundary symplectic form:
\eq{1-5 where}{
\omega = \omega_{\overline\phi\overline\phi}[\overline\phi, 0]\delta\overline\phi\wedge\delta\overline\phi + \dr\omega_b[\overline\phi, \alpha].
}
Like with $H$, the symplectic form separates into a bulk component that we would have found by gauge fixing from the very beginning and a boundary term which encapsulates the entire effect of the gauge symmetry.

We can choose both of these terms to be $\delta$ closed on phase space separately\footnote{To see this, suppose to the contrary that the variation of the bulk term in \eqref{1-5 where} is not $\delta$ closed but varies to something spacetime exact, $\delta(\omega_{\overline\phi\overline\phi}[\overline\phi, 0]\delta\overline\phi\wedge\delta\overline\phi) = \dr\omega_M$, which then cancels against part of $\dr\delta\omega_b$. This $\omega_M$ cannot contain any dependence on $\alpha$ or $\delta\alpha$, so if a cancellation occurs it's sufficient to consider $\alpha = \delta\alpha = 0$. But our use of theorem \ref{Wald Theorem} implies $\omega_b = 0$ in this case, so no cancellation can occur, requiring $\dr\omega_M = 0$, and the terms in \eqref{1-5 where} are separately closed.}. It then follows that we can find separate potentials and the symplectic potential current $\mathcal{A}$ breaks up into two terms,
\eq{1-6 where}{
\mathcal{A}[\overline\phi, \alpha] = \mathcal{A}_M[\overline\phi] + \dr\mathcal{A}_{\p M}[\overline\phi, \alpha].
}

Our arguments about the Hamiltonian and symplectic form together imply that the action must also break up into two terms, one supported on the bulk obtained from gauge fixing, and one supported only on the boundary\footnote{\label{orientation footnote}One should be careful to take the orientation of $\p\Sigma$ in these integrals to be the one induced by writing $\Vol(M) = \tau\wedge n\wedge\Vol(\p\Sigma)$ where $\tau$ is the normal form to the slices $\Sigma$ and $n$ is the (outward) normal form to $\Gamma$. This is a convention consistent with Stokes theorem, see \cite{Harlow:2019yfa} for more details.}:
\eq{1-7 where}{
S = \int\dr t\int_\Sigma\Big( i_V \mathcal{A}_M[\overline\phi] - J_t[\overline \phi, 0] \Big) + \int\dr t\int_{\p \Sigma}\Big( i_V \mathcal{A}_{\p M}[\overline\phi, \alpha] - Q_t[\overline\phi, \alpha]\Big)
}
where $V$ generates the path the action is evaluated on. So all gauge theories will reduce to a bulk component which is gauge fixed, and a boundary term which contains all the dynamics of the boundary gauge modes. Importantly, this boundary action may contain interactions not only among the boundary fields $\alpha$, but also with the boundary values of the bulk gauge fixed fields $\overline\phi$.

\subsection{Discussion}\label{Where Discussion}

The result \eqref{1-7 where} establishes boundary contributions to the action as a generic feature of gauge theory independent of whether we are able to solve constraints and directly reduce the action to the boundary as in 3d CS theory.  But the nature of this boundary action varies from case to case, the main controlling factor being the structure of the asymptotic symmetry group.   For example, Einstein gravity in AdS$_{D>3}$ with standard asymptotically AdS boundary conditions has a finite dimensional asymptotic symmetry given by the SO$(D-1,2)$ isometry group of the global AdS vacuum.  Thus instead of a boundary field theory we have a boundary quantum mechanics, with one quantum mechanical degree of freedom corresponding to each generator.  

To illustrate this  we  consider the even simpler case of pure Maxwell theory in asymptotically flat space.  The phase space variables are $(\vec{A}, \vec{E})$ subject to the Gauss law constraint $\vec{\nabla} \cdot \vec{E}=0$, and we impose that these vanish at spatial infinity. The symplectic form is $\Omega \sim \int_\Sigma \delta \vec{E}\cdot \wedge \delta \vec{A}$.  This symplectic form together with the boundary conditions require that the gauge parameters $\alpha(\vec{x})$ become spatially constant at infinity.    Writing $\vec{A} = \vec{\overline{A}}+ \vec{\nabla} \alpha$, we have 
\eq{ssa}{ \Omega \sim \int_{\p \Sigma} \vec{n}\cdot \delta \vec{E} \wedge \delta \alpha  \sim \delta Q \wedge \delta \alpha   }
where $Q$ is the total electric charge.  As the Hamiltonian has no dependence on $\alpha$, the boundary action is simply $S_\alpha = \int\! dt Q \dot{\alpha}$, whose equation of motion is simply charge conservation.  This boundary contribution is actually familiar in another guise.  In particular, consider the Euclidean theory with periodic imaginary time, $t\cong t+\beta$.  Since $Q$ is constant on-shell we have $S_\alpha= Q \int\! dt \dot{\alpha} \equiv  \beta \mu Q $, which identifies the boundary mode $\alpha$ as being proportional to the chemical potential.  

Turning now to the case of an infinite dimensional asymptotic symmetry group, here  we do expect to get a boundary field theory.  This boundary field theory may either be a free theory, as in $U(1)$ CS theory, or interacting, as in non-Abelian CS theory or 3D gravity.   In general, it's clear that the existence of interactions is closely tied to the bulk theory having a non-Abelian asymptotic symmetry group.

We may also ask more practically how to compute \eqref{1-7 where}. In principle, once we go to gauge orbit variables all stages of the calculation here are algorithmic as reviewed in Appendix \ref{Ap identically closed forms}. But to use the gauge orbit variables we would first need to classify the gauge-inequivalent solutions. Since we are working in a canonical formulation it would be sufficient to find all gauge-inequivalent initial data, but that would still mean solving the constraints of the theory explicitly.

Instead, one could could imagine taking $\overline\phi$ to be some perhaps incomplete class of initial data solving the constraints and consider the dynamics of this subspace of the full phase space. For example by considering a special subspace such as a moduli space of black hole solutions.

The logical extreme of these ideas would be to take $\overline\phi$ to be a single solution, so we are looking at the gauge orbits about a background configuration. In this case $\delta\overline\phi = 0$ and $A_M = 0$ so the entire symplectic form lives on the boundary. This simplification allows more efficient computational methods than the gauge orbit variable strategy described above. Three such methods are described in the next section.

\section{Computing Gauge Orbit Actions}\label{Computing Gauge Orbit Actions}
\label{compute}

As discussed in Section \ref{Where Discussion}, we may have reason to consider only certain sectors of a theory. The particular case of interest here is where we consider a single gauge fixed configuration $\overline\phi$ and orbits around it. This means $\delta\overline\phi = 0$ and $\mathcal{A}_M = 0$. The remainder of the bulk term in the action \eqref{1-7 where} then integrates to some constant and we are left with only the boundary contribution to the action and more efficient methods of calculation are available to us. This section is concerned with describing three such methods.

The first is less generically applicable, but very efficient when it applies as it directly computes $\mathcal{A}_{\p M}$ instead of the boundary symplectic form $\omega_b$. The other two methods are concerned with computing $\omega_b$ and are based on the existence of a particular 1-form valued vector field on phase space.

It's important to note that as $\delta\overline\phi = 0$ suggests, these methods treat $\overline\phi$ as a background field so any components $\mathcal{A}_{\p M}$ might have had in the $\delta\overline\phi$ directions will not be captured by these computations.

\subsection{The momentum method}
\label{momsec}

For all the 3D gravity examples considered in this work there exists a very efficient method for deducing the boundary action \cite{Ebert:2022cle}.  We describe it here in general terms.

Supposing that expressions for boundary charges $H_\xi$ are known, to write down the boundary action we require knowledge of  the boundary symplectic potential $\Theta$, defined via $\Omega = \delta \Theta$.   A direct approach to obtaining it is to first extract $\Omega$ via the relation $i_{V_\xi}\Omega =-\delta H_\xi$, and then  compute $\Theta$ by solving $\delta \Theta = \Omega$.    We now discuss how, under suitable assumptions, we can write down the solution for $\Theta$ directly, bypassing the laborious procedure just mentioned.  This method was used in the case of cutoff AdS$_3$ gravity \cite{Ebert:2022cle}. 

We consider a boundary with a single spatial dimension with coordinate $x$, which may live either on the circle or line.   The boundary field theory variables are written as $(\Phi, \Psi)$, where $\Psi $ could stand for a collection of fields.  On the circle or line, these fields are taken to obey periodic boundary conditions or vanish at infinity respectively.   Under phase space vector fields $V_\xi$, which we can think of as reparametrizations of $x$, the fields transform as
\eq{eea}{ \delta_\xi \Phi &= i_{V_\xi}\delta \Phi = \xi + \Phi' \xi \cr
\delta_\xi \Psi &= i_{V_\xi}\delta \Psi =  \Psi' \xi }
where $' = \p_x$.  The inhomogeneous term in  $\delta_\xi \Phi$ corresponds to $\Phi$ being the field associated with $x$-reparametrizations.   Now suppose that the charge corresponding to constant $\xi$ (we take $\xi=1$ and denote this charge by $P$) is constrained to take the form 
\eq{eeb}{ P= \int \! \dr x \left[ \kappa_{\Phi\Phi} \Phi'^2 +\kappa_{\Phi\Psi} \Phi' \Psi'+ \kappa_{\Psi\Psi} \Psi'^2+ P'_\Phi \Phi' + P'_\Psi \Psi'  \right]  }
where the $\kappa$'s are  constant and  the functions $(P_\Phi,P_\Psi)$ are local functions of $(\Phi', \Phi'', \ldots ;  \Psi', \Psi'', \ldots)$, i.e. do not depend on undifferentiated fields.   The basic result is that up to a $\delta$-exact term, the unique $\Theta$ which solves 
\eq{eeba}{ i_{V_\xi} \Omega =-\delta P~,\quad \xi =1  }
is given by 
\eq{eec}{ \Theta = \int\! \dr x \left[ \kappa_{\Phi\Phi} \Phi' \delta \Phi+   \kappa_{\Phi\Psi} \Phi' \delta \Psi+ \kappa_{\Psi\Psi} \Psi' \delta \Psi+ P'_\Phi \delta \Phi + P'_\Psi \delta \Psi\right]~.}
To prove this, we first note that it is straightforward to verify that \rf{eeba} is satisfied, so only the question of uniqueness remains.   To analyze uniqueness we consider a correction to the symplectic form, 
\eq{eed}{\Delta \Omega = \int\! \dr x \left[ \delta X'_\Phi \wedge\delta \Phi+\delta X'_\Psi \wedge\delta \Psi \right]~, }
where the functions $(X_\Phi,X_\Psi)$ are constrained to obey the same general properties of $(P_\Phi,P_\Psi)$.
We compute
\eq{eee}{ i_{V_\xi}\Delta \Omega &=\delta  \int\! \dr x \left[ X'_\Phi \Phi' + X'_\Psi \Psi' \right] \cr
 &=-\delta  \int\! \dr x \left[ X''_\Phi \Phi + X''_\Psi \Psi \right]  }
We need this to vanish in order not to spoil \rf{eeba}.  Vanishing of  the integrand in the second line  of \rf{eee}  requires $X''_\Phi= X''_\Psi=0$, since the two terms cannot cancel each other under the assumed form of $(X_\Phi,X_\Psi)$.   However, this implies that $(X'_\Phi,X'_\Psi)$ are constants, which implies that $\delta X'_\Phi= \delta X'_\Psi=0$, so that $\Delta \Omega=0$.   The remaining possibility is that the integrand in \rf{eee} is a total derivative.  This requires $X'_\Phi= \frac{\delta F}{\delta \Phi}$ and $X'_\Psi= \frac{\delta F}{\delta \Psi}$  for some $F$.
However, this leads to $\Delta \Omega = \int\! \dr x \delta^2F=0 $, so we again get no contribution.

The upshot is that just from consideration of $P$ we are led to a unique result for $\Omega$ and an explicit result for its potential $\Theta$.   It follows, assuming that our underlying theory is consistent, that this $\Omega$ will solve $i_{V_\xi} \Omega = -\delta H_\xi$ for all  large gauge transformations $V_\xi$, as can of course be checked in specific examples. 
 
We conclude this section with comments on the assumed structure \rf{eeb}, which follows from a particular gauge invariance.  To make the discussion concrete we consider the boundary theory in the context of pure gravity in AdS$_3$.   Setting $\kappa=0$ corresponds to the orbit built on a pure AdS$_3$ background.   This background is invariant under isometries that act as translations, boosts, and dilatations on the boundary coordinates.  Taking the functions $(\Phi,\Psi)$ to be general linear functions of $x$ corresponds to acting on the background by one of these isometries; since this does nothing to the background such functions are pure gauge, and so all charges must vanish for such functions.  This implies that only derivatives of such functions can appear in $P$, and that each term must contain at least one second derivative.   This leads to \rf{eeb}, possibly after integrating by parts.  The same comments apply to $\Delta \Omega$.  Turning on $\kappa$, which  corresponds to  a nonzero mass AdS$_3$ background, breaks some of the isometries, allowing the quadratic terms like $\kappa_{\Phi\Phi} \Phi'^2$  to appear.   In fact, a more general expression like $\kappa_{\Phi\Phi} \Phi'^n$ for some $n>2$ might also be anticipated, but only the $n=2$ case will arise in our examples.  This is important, since the logic leading to \rf{eec} no longer holds for the $n>2$ case.  In fact, given  the simple form of the $\kappa$ dependent part of that charges  will arise, it's easy to write down the corresponding contribution to the boundary action by inspection.

\subsection{The transfer field}

Consider a variation of our fields in the gauge orbit variables $\phi = T_\alpha[\overline\phi]$, as in Section \ref{Where Boundary Actions Come From}. Since we are taking $\overline\phi$ to be fixed, the variation can only change $\alpha$. But since $\phi$ is constructed by applying a gauge group element to the configuration $\overline\phi$, this variation must be equivalent to the action of some Lie algebra element on $\phi$.

The situation is essentially identical to the story which comes up when studying spontaneous symmetry breaking, in particular the broken part of the symmetry group's non-linear action on Nambu-Goldstone bosons  when studying spontaneous symmetry breaking, see for example \cite{Coleman:1969sm,Callan:1969sn}. The idea is that $\delta\alpha$ defines a Lie algebra element at the origin of the gauge group while $\overline\phi$ has been transported to $g$. So to perform the variation of $\phi$ we need to transport $\delta\alpha$ to the tangent space at $g$. Schematically, we can think of this as inserting the identity in the action on $\overline\phi$ to write
\eq{1 1formV}{
\delta\phi = (\delta T_g)[\overline\phi] = (\delta T_g T^{-1}_g)[\phi]
}
so the variation of $\phi$ is equivalent to the action of this Lie algebra element valued as a 1-form on phase space.

Since Lie algebra elements are the generators of group transformations, there must exist some $W$ which implements the action of this Lie algebra element. Because the algebra element is valued as a 1-form on phase space, $W$ must be a $(1, 1)$ tensor on phase space, which we think of as a 1-form valued vector field and refer to as the transfer field. This gives $W$ the very special defining property
\eq{2 1formV}{
\delta\phi = i_W\delta\phi.
}

While the property \eqref{2 1formV} is ultimately the formal property we will exploit to compute the boundary symplectic form, we will also need to explicitly compute the transfer field. So it will be useful to see how the rather abstract discussion above can be realized in some simple examples.

The simplest example is of a $U(1)$ connection $A = \overline A + \dr\alpha$. Here we have explicitly
\eq{3 1formV}{
\delta A = \dr\delta\alpha = i_{V_{\delta\alpha}}\delta A
}
so $W = V_{\delta\alpha}$ implements the gauge transformation with gauge parameter $\delta\alpha$, as might be expected for an Abelian group where translating $\delta\alpha$ is trivial.

The next simplest case is where $A$ is a connection for some non-Abelian group. Now the gauge orbit is $A = g^{-1}\overline A g + g^{-1}\dr g$. Here $g = g(\alpha)$ is defined by the exponential map. For simplicity we could take $g = \exp(\alpha)$ with $\alpha$ valued in the Lie algebra. We may now write
\eq{4 1formV}{
\delta A = [A, g^{-1}\delta g] + \dr(g^{-1}\delta g) = i_{V_{g^{-1}\delta g}}\delta A
}
so $W = V_{g^{-1}\delta g}$ generates a gauge transformation with 1-form valued gauge parameter $g^{-1}\delta g$. We also note that this examples gives a concrete realization of the schematic manipulation \eqref{1 1formV}.

As a final example of $W$ for this section, we consider diffeomorphisms. Take the diffeomorphism to be $y^\alpha = f^\alpha(x)$ and consider for simplicity the diffeomorphism orbit of a scalar field $\phi(x) = \overline\phi(f(x))$. Taking the variation of this scalar field,
\eq{5 1formV}{
\delta \phi &= \delta f^\alpha(x)\frac{\p\overline\phi(y)}{\p y^\alpha}\Big|_{y = f(x)} \cr
&= \delta f^\alpha(x)\frac{\p x^\mu}{\p y^\alpha}\Big|_{y = f(x)}\frac{\p \overline\phi(f(x))}{\p x^\mu} \cr
&= \eL_\xi\phi
}
so the variation is equivalent to the action of the Lie derivative with respect to the 1-form valued (spacetime) vector field\footnote{In practical calculations, it is much simpler to move the Jacobian factor to the left and compute the variation of $f$ by a diffeomorphism, then defining $\xi$ by inverting the Jacobian. This is the approach taken in the examples of Sections \ref{AdS GR} and \ref{New AdS GR}.}
\eq{6 1formV}{
\xi^\mu = \frac{\p (f^{-1})^\mu}{\p y^\alpha}\Big|_{f(x)}\delta f^\alpha(x).
}
Hence $W = V_\xi$. This fact was also noted in \cite{Harlow:2019yfa} where it was used to circumvent some technical difficulties in JT gravity.

With these examples in mind, there is one additional property of $W$ which will be important going forward. Denote by $w$ the Lie algebra element whose action $W$ generates. In our examples we have $w = \delta\alpha$ for the $U(1)$ gauge group, $w = g^{-1}\delta g$ for the non-Abelian gauge group, and for diffeomorphisms we have \eqref{6 1formV}. 

Importantly, this $w$ is some functional of $\delta\alpha$. Equally as important, this functional need not be invertible as can be seen in the non-Abelian example\footnote{Invertibility fails for the same reason $A = g^{-1}\dr g$ is not invertible for $\dr\alpha$: if $A$ is not flat no such $\dr\alpha$ exists.}. This will be important going forward because to apply Theorem \ref{Wald Theorem} we need free functions. In our setup $\delta\alpha$ are free, but $w$ may or may not be, depending on the details of the gauge group. For example, the relation between $\delta f$ and $\xi$ in \eqref{6 1formV} remains invertible despite diffeomorphisms being non-Abelian.

While the relation between $w$ and $\delta\alpha$ is not invertible, it is injective, and hence is invertible on the image of all $\delta\alpha$. This means $\delta w$ may always be expressed in terms of only $w$ again. We can see this in our examples. The $U(1)$ case is uninteresting because $\delta w = 0$, but in the non-Abelian case we find
\eq{7 1formV}{
\delta w = \delta(g^{-1}\delta g) = - (g^{-1}\delta g)\wedge(g^{-1}\delta g) = -w\wedge w,
}
so the general invertibility of $w(\delta\alpha)$ is unimportant to this rewriting of $\delta w$, as claimed.

The example of diffeomorphisms is slightly more complicated, but nonetheless can be worked out to find
\eq{8 1formV}{
\delta \xi^\mu &= \xi^\nu\wedge\nabla_\nu\xi^\mu,
}
assuming we are working with a torsionless connection so the affine part of the covariant derivative does not contribute to this expression.

\subsection{The transfer field method}\label{iWTheta}

Now that we have an understanding of the transfer field $W$, we can give a technique for computing the boundary symplectic form. From the definition \eqref{4-2 canonical}, it would clearly be sufficient to show that the bulk $\delta\theta$ term is a total derivative. We may use the transfer field to write
\eq{9 1formV}{
\theta = i_W\theta
}
since in every term of $\theta$, the 1-form factor can be replaced as in \eqref{2 1formV}.

Since $W$ generates the action of the Lie algebra element $w$, it follows that
\eq{10 1formV}{
\theta = i_W\theta = J_{w} + k_w
}
where $J_{w}$ is the Noether current \eqref{7-1 canonical} associated to the gauge transformation generated by $w$ and $k_w$ is defined by \eqref{7 canonical}.

But as discussed in Section \ref{Sec: cov ps formalism}, $J_w = \dr Q_w$ and hence
\eq{11 1formV}{
\omega = \delta\theta - \dr \delta B = \delta k_w + \dr \delta(Q_w - B).
}
So to find $\omega_b$ it is  sufficient to find a potential for $\delta k_w$ by solving $  \delta k_w= \dr \tilde k_w$.

Since by \eqref{6 canonical} we have $\dr\omega = 0$ on-shell, we must also have $\dr\delta k_w = 0$\footnote{We could also argue more directly that $\delta\dr k_w = \delta i_W\delta L = \delta^2 L = 0$ since $i_W\delta L = \delta L$ by \eqref{2 1formV}.}. Now $\delta k_w$ depends on the free functions\footnote{In general, the imposition of boundary conditions will restrict what functions $\alpha$ we allow ourselves to consider near the boundary. However, the only place the boundary conditions enter in these arguments is through $B$, and in particular the closure of $\delta k_w$ requires no reference to boundary conditions. So we should imagine performing these manipulations before imposing any boundary conditions which would then restrict the allowed $\delta\alpha$.} $\delta\alpha$ so we may apply theorem \ref{Wald Theorem} to conclude the existence of a locally constructed $\tilde k_w$.

It should be commented that while $\delta\alpha$, and the fact that $\delta\alpha = 0$ corresponds to the identity, ensures the existence of $\tilde k_w$, it would be more computationally efficient to use $w$ as the free function whenever possible. This can be done when the relation between $w$ and $\delta\alpha$ is invertible, as discussed in the previous section. The invertibility in the case of diffeomorphisms will be leveraged in Section \ref{AdS GR}.

Hence we have found
\eq{12 1formV}{
\omega_b = \tilde k_w + \delta(Q_w - B)
}
which has the advantage of cleanly separating the boundary condition dependence of $\omega_b$ into the $\delta B$ term since $\tilde k_w$ and $Q_w$ depend only on the Lagrangian.

This technique is straightforward to apply to $U(1)$ CS theory on $D\times\mathbb{R}$ with chiral boundary conditions. We recall for this theory \eqref{afa-1}, note $k_\lambda = k\lambda \dr A$, and consider a solution $\overline A$ to the equations of motion, so $\dr \overline A = 0$. The gauge orbit around this solution is then $A = \overline A + \dr\alpha$. As discussed in the previous section we have $w = \delta\alpha$.

Since $k_\lambda = 0$ on-shell in this theory, we have $\tilde k_w = 0$. Then following \eqref{12 1formV}, the boundary symplectic form must be
\eq{15 1formV}{
\omega_b = \delta Q_w = k\delta A \wedge \delta\alpha
}
where we have used $B = 0$. This integrates to
\eq{16 1formV}{
\Omega = k\int_0^{2\pi}\dr\phi \,\p_\phi\delta\alpha \wedge \delta\alpha
}
and produces the usual Kac-Moody algebra \eqref{af} expected for $U(1)$ CS theory on the disk. Using the Hamiltonian \eqref{ae} and the phase space action \eqref{1-7 where} we reproduce the expected boundary action \eqref{ad}.

\subsection{Computing \texorpdfstring{$\Omega$}{Omega} from the Noether charges}\label{Noether Charge}

In the previous section we were able to leverage the transfer field to write \eqref{9 1formV}. It's simple to see that this generalizes to any 1-form on the gauge orbit, not just $\theta$. With some additional though we can go beyond $1$-forms  and obtain a similar result for any $p$-form on the orbit.

To see how this works, consider $i_W(\delta\phi^a \wedge \delta\phi^b)$ where we have restored the indices $a$ and $b$ labeling our fields. Computing the contraction we find
\eq{1-1 ivomega}{
i_W (\delta\phi^a \wedge\delta\phi^b) &= (i_W\delta\phi^a)\wedge \delta \phi^b + \delta\phi^a \wedge (i_W\delta\phi^b) \cr
&= 2 \delta\phi^a \wedge \delta \phi^b
}
where in the second line we have used the defining relation \eqref{2 1formV} for $W$. It's important to keep track of the sign on the second term because $W$ is 1-form valued and must be commuted past the first factor.

With this sign understood, the generalization to a form of any degree is immediate: we simply march through the factors in the wedge product to perform the contraction and then use the definition of $W$ to rewrite the contraction back in terms of the original form. Hence, for any $p$-form $Z$ on phase space we have the generalized identity
\eq{1-2 ivomega}{
i_WZ = pZ.
}

We can leverage this to very efficiently compute the boundary symplectic form $\omega_b$ on the gauge orbit directly from the Noether charges. The key is to observe that the contraction on an arbitrary 2-form may be rewritten into the form
\eq{1-3 ivomega}{
i_W(\delta\phi^a\wedge\delta\phi^b) = -\left[ \delta\phi^b \wedge (i_W\delta\phi^a) - \delta\phi^a \wedge (i_W\delta\phi^b) \right].
}
That is, if we make sure to commute the contracted part to the right of the expression, we obtain the negative of the expression we would have found if $W$ was not valued as a 1-form.

It follows that if we compute $i_W\Omega$ and make sure to commute all the contractions to the right, we must obtain
\eq{1-4 ivomega}{
i_W\Omega = +(\delta H[\tilde w])|_{\tilde w = w}
}
since we have already argued that $W$ generates the action of a Lie algebra element. Note that on the right hand side we compute the variation of $H$ before evaluating at the gauge parameter $w$. This is because $w$ enters through the contraction on the right and so clearly cannot have a variation applied to it.

Combining this with our general observation \eqref{1-2 ivomega} it follows that the symplectic form on the gauge orbit is given by
\eq{1-5 ivomega}{
\Omega = \frac{1}{2}(\delta H[\tilde w])|_{\tilde w = w}
}
where the expression on the right should be understood as placing all factors of $w$ to the right of any variations, and the variation taken treating $\tilde w$ as an arbitrary, but not 1-form valued, Lie algebra element. One must be cautious not to interpret \eqref{1-5 ivomega} as meaning $\Theta = \frac{1}{2}H[w]$. The order in which the variation and the evaluation at $\tilde w = w$ occur are important for this result, as will be seen our examples in Section \ref{Examples}.

We may again revisit our $U(1)$ CS example to demonstrate this approach. Indeed, looking at \eqref{15 1formV} we can already see the basic structure of \eqref{1-5 ivomega} since $B = 0$ with our chiral boundary conditions. But going through the details to make sure the coefficients come out correct, particularly since \eqref{15 1formV} uses the Noether current and not the complete Noether charge, we start with the charges \eqref{ae}. Then \eqref{1-5 ivomega} tells us to vary the charge, treating $\tilde w$ as any large gauge parameter in the theory, which generally may mean it's state-independent or some state-dependent function of some other parameters determining the large gauge transformations. Here there are no such complications as $\tilde w$ is state-independent and we find
\eq{1-7 ivomega}{
\frac{1}{2}\delta H[\tilde w] = k\int_0^{2\pi}\tilde w\delta A_\phi\dr\phi.
}
The final step in computing \eqref{1-5 ivomega} is to evaluate this charge at $\tilde w = w \equiv \delta\alpha$. But when we do so, we must make sure that we first move all $\tilde w$ factors to the right of $\delta A_\phi$. Doing so,
\eq{1-8 ivomega}{
\Omega = k\int_0^{2\pi}\dr\phi \delta A_\phi \wedge \delta\alpha = k\int_0^{2\pi}\dr\phi \p_\phi\delta\alpha \wedge \delta\alpha
}
which again matches the expected Kac-Moody expression \eqref{ad}.

\subsection{Examples}\label{Examples}

Here we apply our methods to three examples. First we consider $U(1)$ CS theory in $D = 5$. This theory, with our chosen boundary conditions, is only marginally more complex than the example of $U(1)$ CS theory in $D = 3$ that we have been considering. It is also simple enough that we are able to implement the manipulations in Section \ref{Where Boundary Actions Come From} to find \eqref{1-5 where} with non-gauge directions included.

In Sections \ref{AdS GR} and \ref{New AdS GR} we consider Einstein-Hilbert gravity in $D = 3$ with two different sets of boundary conditions. First we consider standard asymptotically AdS$_3$ boundary conditions and then the boundary conditions described in \cite{Compere:2013bya}. The latter are designed to produce a theory similar to what we will see in topologically massive gravity, considered in Section \ref{Application to Topologically Massive Gravity}.

Finally, we mention that the example of $SU(2)$ CS theory in $D = 3$ is considered in Appendix \ref{Non-Abelian CS}. There we also show that the results of our methods match the standard results expected from the CS/WZW correspondence.

\subsubsection{U(1) CS in $D = 5$}

As a first example which isn't quite as trivial as $U(1)$ CS in $D = 3$ we consider $U(1)$ CS in $D = 5$.    Since we work near the boundary the details of the bulk geometry are irrelevant.  We fix the boundary geometry, though the details of this geometry will not be paramount to most of our manipulations, to be $\mathbb{R}\times S^1\times S^2$. On the boundary we choose coordinates $x^i$, $i = 1,2,3,4$ with $x^1$ the coordinate on the $\mathbb{R}$, which we think of as time, $x^2$ the angular coordinate on $S^1$, and $x^3,x^4$ some coordinates on the $S^2$. To make contact with the $D = 3$ CS theory, it will sometimes be useful to write $t \equiv x^1$ and $\phi \equiv x^2$.

For this example we choose our boundary conditions to fix $A_3$ and $A_4$ to be arbitrary functions of $x^3$ and $x^4$ on the boundary while we fix $A_1 = A_2$ to be an arbitrary function of $x^1$ and $x^2$. This will allow us to draw parallels to $U(1)$ CS in $D = 3$ on the cylinder in the final result. It is useful to note that with these boundary conditions the only non-zero boundary components of $F$ are $F_{12}$, which depends dynamically on $A_1$ (and hence $A_2$), and $F_{34}$ which is a fixed function of only $x^3$ and $x^4$.

The Lagrangian for this theory is
\eq{1 5d CS}{
L = A \wedge F \wedge F
}
which varies to produce
\eq{2 5d CS}{
\delta L = 3 F \wedge F \wedge \delta A - 2 \dr(F \wedge A \wedge \delta A).
 }
Thus the equations of motion, $F\wedge F = 0$, do not imply that all solutions are flat connections. Considering the condition \eqref{4 canonical}, we see that our chosen boundary conditions allow us to choose $\ell = B = 0$.

Since $B = 0$, the full symplectic current for the theory is given by
\eq{3 5d CS}{
\omega = - 2 \delta(F \wedge A) \wedge \delta A.
}
Now, this theory is sufficiently simple that we are able to explicitly carry out the gauge-orbit described in Section \ref{Where Boundary Actions Come From}. We suppose $\overline A$ is some solution to the equations of motion and we consider the gauge orbits around this configuration, $A = \overline A + \dr\alpha$. It's straightforward to find
\eq{4 5d CS}{
\omega = -2 \delta(\overline F \wedge \overline A) \wedge \delta\overline A - 2\dr[ \delta(\alpha\overline F)\wedge \dr\delta\alpha + \delta(\alpha \overline F) \wedge \delta\overline A - \delta(\overline F \wedge \overline A) \wedge \delta\alpha ]
}
so we realize the expected split into the gauge-fixed part and the boundary part depending on the large gauge transformations parametrizing the orbit. Here we can also see that the boundary component of the symplectic form supports components which mix variations in the gauge-fixed configuration with variations along the orbit. Indeed, there is even a component which involves no variations of $\alpha$.

But here we are interested only in the components along the orbit directions, so we take $\delta \overline A = 0$ to find
\eq{5 5d CS}{
\omega_b = -2 \overline F  \delta\alpha \wedge \dr\delta\alpha.
}
Integrating this over the boundary and using our boundary conditions the full symplectic form on the orbit is given by
\eq{6 5d CS}{
\Omega &= -2 \int_{\p\Sigma}\overline F_{34}\delta\alpha \wedge \p_2\delta\alpha \dr x^2 \wedge \dr x^3 \wedge \dr x^4 \cr
&= 2\left(\int_{S^2}\overline F\right)\int_{0}^{2\pi}\p_\phi \delta\alpha \wedge \delta\alpha\dr\phi.
}
Because we have chosen our boundary conditions to factor the $S^2$ from the $\mathbb{R}\times S^1$, we obtain the same Kac-Moody symplectic form as would be expected from the $U(1)$ CS theory on the cylinder, but now with an effective level set by our boundary conditions via the magnetic flux through the $S^2$.

Using that the Hamiltonian generating time evolution is
\eq{7 5d CS}{
H_t &= 2\int_{\p\Sigma} \overline F_{34}A_2^2 \dr x^2 \wedge \dr x^3 \wedge \dr x^4 \cr
&= 2\left(\int_{S^2}\overline F\right)\int_0^{2\pi}(\overline A_\phi + \p_\phi\alpha)^2\dr\phi.
}
Constructing the phase space action via \eqref{1-7 where} we thus find
\eq{8 5d CS}{
S = 2\left(\int_{S^2}\overline F\right)\int\left[ \p_\phi\alpha\left(\p_t\alpha - \p_\phi\alpha - 2 \overline A_\phi\right) - \overline A_\phi^2 \right] \dr t\dr \phi.
}
Since we have restricted ourselves to the gauge orbit, $\alpha$ is our only dynamical variable in this action and, in particular, this means the final $\overline A_\phi^2$ term can be dropped as an additive constant to the action. In this example it's possible to observe explicitly from \eqref{4 5d CS} that, had we not restricted to the gauge orbit of some fixed $\overline A$, the $\overline A_\phi$ terms here would represent an explicit coupling between the bulk, $\overline A$, and boundary, $\alpha$, degrees of freedom.

This result can, of course, also be obtained by the techniques introduced in the previous sections. Taking first the approach of Section \ref{iWTheta} we note that
\eq{9 5d CS}{
i_{V_\lambda}\delta L = \dr \lambda \wedge F \wedge F = \dr(\lambda F \wedge F)
}
so $k_\lambda$ vanishes on-shell and will make no contribution. For the other ingredient, we calculate
\eq{10 5d CS}{
J_\lambda = i_{V_\lambda}\theta - k_\lambda = 2\dr(\lambda A \wedge F).
}

Thus, since $B = 0$ with our chosen boundary conditions, we find from \eqref{15 1formV}
\eq{11 5d CS}{
\omega_b = 2\delta(A \wedge F) \wedge \delta \alpha = 2 \dr\delta\alpha \wedge \overline F \wedge \delta\alpha
}
since $w = \delta\alpha$ for the $U(1)$ orbit, matching \eqref{5 5d CS}. From this point, imposing the boundary conditions to find the true symplectic form \eqref{6 5d CS}, and hence the action \eqref{8 5d CS}, is unchanged.

If we instead took the route of Section \ref{Noether Charge} we would need to compute the full Noether charges
\eq{12 5d CS}{
H[\lambda] = 4 \left(\int_{S^2}\overline F\right)\int_{S^1}\lambda A
}
which requires that we use the boundary conditions to obtain. For example, that we fix $A_3$ and $A_4$ completely on the boundary tells us that $\lambda = \lambda(x^1, x^2)$. The formulation \eqref{1-5 ivomega} now tells us to write
\eq{13 5d CS}{
\Omega = 2\left(\int_{S^2}\overline F\right)\int\delta A \wedge \delta\alpha.
}
Writing this explicitly in coordinates and with $\delta A = \delta\dr\alpha$ this clearly reproduces \eqref{6 5d CS} and the rest of the boundary action story follows.

\subsubsection{Asymptotically AdS$_3$ Einstein-Hilbert gravity}\label{AdS GR}

As a second example, we can derive the Alekseev-Shatashvili  symplectic form \cite{Alekseev:1988ce} for asymptotically AdS$_3$ Einstein-Hilbert gravity. All three approaches can be worked out in this example. We start by recalling some facts this theory.

The action and canonical 1-form  for this theory are given by
\eq{1 examples}{
L = \frac{1}{16\pi G}\sqrt{-g}(R + 2)\dr^3 x,\ \ \ \ \theta = \frac{1}{16\pi G}\sqrt{-g}(\nabla^\nu\delta g_{\lambda\nu} - g^{\mu\nu}\nabla_\lambda\delta g_{\mu\nu})(\dr^2x)^\lambda.
}
Where expressions are more conventionally expressed in terms of the Brown-Henneaux central charge we write $c = 3/2G$. We will also need later that
\eq{2 examples}{
k_\xi = i_\xi L = -\frac{1}{4\pi G}\sqrt{-g}\xi^\mu (\dr^2 x)_\mu
}
where we have used the equations of motion to simplify things.

In Fefferman-Graham coordinates the asymptotically AdS$_3$ solutions to the equations of motion are given by the Ba\~nados metrics \cite{Banados:1998gg}
\eq{3 examples}{
\dr s^2 = \frac{\dr\rho^2}{4\rho^2} + \frac{1}{\rho}(\dr w + \rho \overline L(\overline w)\dr \overline w)(\dr \overline w + \rho L(w)\dr w)
}
where $w = \phi + t$ and $\overline w = \phi - t$ are convenient coordinates on the boundary and $\rho > 0$ is the radial coordinate such that the boundary is located at $\rho 
= 0$. Note that $\overline w$ is not the complex conjugate of $w$ as we are working in Lorentzian signature. The boundary stress tensor associated to these metrics is given by
\eq{4 examples}{
T_{ww} = - \frac{1}{4G}L,\ \ \ \ T_{\overline w \overline w} = - \frac{1}{4G}\overline L,\ \ \ \ T_{w\overline w} = 0.
}
The asymptotic vector fields which preserve the boundary conditions are
\eq{5 examples}{
\xi^w &= \epsilon(w) - \frac{1}{2}\p^2_{\overline w}\overline \epsilon(\overline w) \rho + \mathcal{O}(\rho^2)\\
\xi^{\overline w} &= \overline\epsilon(\overline w) - \frac{1}{2}\p_w^2\epsilon(w)\rho + \mathcal{O}(\rho^2)\\
\xi^\rho &= \left( \p_w \epsilon(w) + \p_{\overline w}\overline \epsilon(\overline w) \right)\rho+ \mathcal{O}(\rho^0),
}
where $\epsilon(w)$ and $\overline\epsilon(\overline w)$ are free functions labeling the vector field. While these vector fields do not change the conformal metric on the boundary, they do change the components subleading in $\rho$, and hence the stress tensor by
\eq{6 examples}{
i_{V_\xi}\delta T_{ww} &= 2 T_{ww}\epsilon' + T'_{ww}\epsilon + \frac{c}{12}\epsilon'''\\
i_{V_\xi}\delta T_{\overline w\overline w} &= 2T_{\overline w\overline w}\overline \epsilon' + T_{\overline w\overline w}'\overline \epsilon + \frac{c}{12}\overline\epsilon'''.
}
The charges associated with the diffeomorphisms are then easily constructed by
\eq{7 examples}{
H[\xi] = -\frac{1}{2\pi}\int_0^{2\pi}(T_{ww}\epsilon - T_{\overline w\overline w}\overline \epsilon) \dr \phi.
}

Going to the gauge orbit means integrating up the variations \eqref{6 examples} under an infinitesimal diffeomorphism to a finite one. If the finite diffeomorphism on the boundary is given by
\eq{8 examples}{
w' = f(w), \ \ \ \ \overline w' = \overline f(\overline w)
}
then we have
\eq{9 examples}{
T_{ww} = \frac{c}{12}\left(\frac{\kappa}{2}f'^2 + \{f, w\}\right),\ \ \ \ T_{\overline w\overline w} = \frac{c}{12}\left( \frac{\overline\kappa}{2}\overline f'^2 + \{\overline f,\overline w\}\right)
}
where the Schwarzian derivative is
\eq{9-1 examples}{
\{f, w\} = \frac{f'''}{f'} - \frac{3}{2}\frac{f''^2}{f'^2}.
}
In this orbit, the diffeomorphism $f(w) = w$ and $\overline f(\overline w) = \overline w$ evidently produces constant $T_{ww}, T_{\overline w\overline w}$. The $\kappa$ and $\overline\kappa$ parametrize what this constant value is, and thereby parametrize the diffeomorphism inequivalent metrics \eqref{3 examples}. The case $\kappa = \overline\kappa = 1$ produces global AdS.

At $\rho = 0$, \eqref{5 examples} defines a diffeomorphism on the boundary which acts on the $f$ and $\overline f$ as
\eq{10 examples}{
i_{V_\xi}\delta f = f' \epsilon, \ \ \ \ i_{V_\xi}\delta \overline f = \overline f' \overline\epsilon.
}
This can be obtained by composing two diffeomorphisms and we may also verify that this transformation rule is compatible with \eqref{6 examples} and \eqref{9 examples} together. Since we have the momentum written down now as
\eq{9-1-1 examples}{
P = H[\p_\phi] = -\frac{c}{24\pi}\int_0^{2\pi}\dr\phi\left[ \left( \frac{\kappa}{2}f' - \frac{1}{2}\left(\frac{1}{f'}\right)''\right) f' - \left(\frac{\overline\kappa}{2}\overline f' - \frac{1}{2}\left(\frac{1}{\overline f'}\right)''\right)\overline f' \right]
}
where we have used that $\{f, w\} = -\frac{1}{2}\left(\frac{1}{f'}\right)''f'$ up to addition of total derivatives, we may observe that it takes the form \eqref{eeb} required for the momentum method.

It therefore follows that the canonical 1-form is given by
\eq{13-1 examples}{
\Theta = -\frac{c}{24\pi}\int_0^{2\pi}\dr\phi\left[\left( \frac{\kappa}{2}f' - \frac{1}{2}\left(\frac{1}{f'}\right)''\right) \delta f - \left(\frac{\overline\kappa}{2}\overline f' - \frac{1}{2}\left(\frac{1}{\overline f'}\right)''\right)\delta\overline f \right].
}
The variational of this potential can indeed be massaged into the more standard form
\eq{13 examples}{
\Omega = -\frac{c}{48\pi}\int_0^{2\pi}\left[ \left( \frac{\delta f' \wedge \delta f''}{f'^2} - \kappa \delta f\wedge \delta f' \right) - \left(\frac{\delta \overline f' \wedge\delta \overline f''}{\overline f'^2} - \overline\kappa \delta\overline f \wedge\delta\overline f' \right) \right]\dr \phi.
}
of the Alekseev-Shatashvili symplectic form.

Using the canonical 1-form \eqref{13-1 examples} in \eqref{1 where} together with the Hamiltonian $H[\p_t]$ produces the Alekseev-Shatashvili action
\eq{13-2 examples}{
S = - \frac{c}{24\pi}\int\dr^2x\left[ \kappa f' \p_{\overline w}f - \left(\frac{1}{f'}\right)''\p_{\overline w}f - \overline \kappa \overline f' \p_w \overline f - \left(\frac{1}{\overline f'}\right)''\p_w\overline f \right].
}

We may also obtain these results by the technique in Section \ref{Noether Charge}. Since we already know the charges, the only ingredient still needed in \eqref{1-5 ivomega} is the 1-form valued gauge parameter $w$. Since we are working with diffeomorphisms this means we need to work out \eqref{6 1formV}. In this case things are simplified somewhat because we evidently only require the vector field $\xi$ evaluated at $\rho = 0$ since the subleading components of \eqref{5 examples} are not needed to compute the charge $H$.

Now, \eqref{10 examples} gives the variations of $f$ and $\overline f$ under an arbitrary asymptotic diffeomorphism, the 1-form valued vector field with parameters $\epsilon_\delta$ and $\overline \epsilon_\delta$ must satisfy
\eq{11 examples}{
\left(\begin{array}{c}
\delta f\\
\delta\overline f
\end{array}\right) = \left(\begin{array}{cc}
f' & 0\\
0 & \overline f'
\end{array}\right)\left(\begin{array}{c}
\epsilon_\delta\\
\overline \epsilon_\delta
\end{array}\right).
}
This is evidently \eqref{6 1formV} evaluated on the boundary and with the Jacobian factor moved to the other side of the equality.

It now follows that \eqref{1-5 ivomega} gives
\eq{12 examples}{
\Omega &= -\frac{1}{4\pi}\int_0^{2\pi}(\delta T_{ww} \wedge \epsilon_\delta - \delta T_{\overline w\overline w} \wedge \overline \epsilon_\delta)\dr\phi \cr 
&= -\frac{c}{48\pi}\int_0^{2\pi}\left[ \delta\left( \frac{\kappa}{2}f'^2 + \{f, w\} \right) \wedge \frac{\delta f}{f'} - \delta\left( \frac{\overline\kappa}{2}\overline f'^2 + \{\overline f, \overline w\}\right) \wedge \frac{\delta\overline f}{\overline f'} \right]\dr\phi.
}
Integrating by parts this expression can can be shown to equal \eqref{13 examples}.

We can also obtain this result by computing \eqref{12 1formV}. For this computation we will need the potential $Q_\xi$ for the Noether current $J_\xi$ and the potential $\tilde k_\xi$ for $\delta k_\xi$. Much of this discussion is independent of the choice $\Lambda = -1$ and $D = 3$, so we will leave these undetermined until they are actually needed to simplify our expressions. The Noether-Wald charge is well known as the so-called Komar term \cite{Iyer:1994ys, Compere:2018aar},
\eq{14 examples}{
Q_\xi = -\frac{1}{16\pi G}\sqrt{-g}\nabla^\mu\xi^\nu(\dr^{D-2}x)_{\mu\nu}.
}

The computation of $\tilde k_\xi$ is straightforward from \eqref{2 examples} if we use \eqref{8 1formV}. With this we find
\eq{15 examples}{
16\pi G\delta k_\xi &= -\frac{4 \Lambda}{D - 2}\left[ \delta\xi^\mu \sqrt{-g} + \frac{1}{2}\sqrt{-g}g^{\alpha\beta}\delta g_{\alpha\beta} \wedge \xi^\mu \right](\dr^{D-1}x)_\mu \cr
&= \frac{4\Lambda}{D - 2}\sqrt{-g}\nabla_\lambda(\xi^\lambda \wedge \xi^\mu) (\dr^{D-1}x)_\mu.
}
From this we identify
\eq{16 examples}{
\tilde k_\xi = - \frac{1}{16\pi G}\frac{2\Lambda}{D - 2}\sqrt{-g}\xi^\mu \wedge \xi^\nu (\dr^{D-2}x)_{\mu\nu}.
}

The boundary symplectic form is therefore given by \eqref{12 1formV} to find
\eq{17 examples}{
\omega_b &= \delta Q_\xi + \tilde k_\xi \cr
&= -\frac{1}{16\pi G}\sqrt{-g}\left[ \nabla_\lambda \xi^\lambda \wedge \nabla^\mu \xi^\nu + \delta(\nabla^\mu\xi^\nu) + \frac{2\Lambda}{D - 2} \xi^\mu\wedge\xi^\nu \right](\dr^{D - 2}x)_{\mu\nu}.
}
Computing
\eq{18 examples}{
\delta(\nabla^\mu\xi^\nu)(\dr^{D-2}x)_{\mu\nu} = -\left[ \nabla^\lambda\xi^\mu \wedge\nabla_\lambda \xi^\nu - \frac{1}{2}R_{\alpha\beta}^{\ \ \ \mu\nu}\xi^\alpha\wedge\xi^\beta \right] (\dr^{D-2}x)_{\mu\nu}
}
the boundary symplectic form becomes
\eq{19 examples}{
\omega_b = -\frac{1}{16\pi G}\sqrt{-g}\left[ \nabla_\lambda\xi^\lambda \wedge \nabla^\mu\xi^\nu - \nabla^\lambda\xi^\mu \wedge\nabla_\lambda \xi^\nu - \frac{1}{2}R_{\alpha\beta}^{\ \ \ \mu\nu}\xi^\alpha\wedge\xi^\beta + \frac{2\Lambda}{D - 2} \xi^\mu \wedge \xi^\nu \right](\dr^{D - 2}x)_{\mu\nu}.
}

If we now specialize to $D = 3$ where
\eq{20 examples}{
R_{\alpha\beta}^{\ \ \ \mu\nu} = \delta^\mu_\alpha R^\nu_\beta - \delta^\nu_\alpha R^\mu_\beta - \delta^\mu_\beta R^\nu_\alpha + \delta^\nu_\beta R^\mu_\alpha - \frac{1}{2}R(\delta_\alpha^\mu\delta^\nu_\beta - \delta^\nu_\alpha\delta^\mu_\beta)
}
and use the equations of motion to write
\eq{21 examples}{
R_{\mu\nu} = 2\Lambda g_{\mu\nu},\ \ \ \ R = 6\Lambda,
}
\eqref{19 examples} can be simplified to
\eq{22 examples}{
\omega_b = - \frac{c}{24\pi}\sqrt{-g}\Big[ \nabla_\lambda \xi^\lambda \wedge \nabla^\mu\xi^\nu - \nabla^\lambda \xi^\mu \wedge \nabla_\lambda \xi^\nu + \Lambda \xi^\mu \wedge \xi^\nu \Big] (\dr^1x)_{\mu\nu}.
}
It's straightforward to evaluate \eqref{22 examples} on the metric \eqref{3 examples} on the computer. We restrict the vector field $\xi$ to take the form \eqref{5 examples}, where now $\epsilon$ and $\overline \epsilon$ should be understood as the 1-form valued $\epsilon_\delta$ and $\overline\epsilon_\delta$ we used earlier and introduced in \eqref{11 examples}. Doing so we find the $\phi = \frac{1}{2}(w + \overline w)$ component of \eqref{22 examples} to be
\eq{23 examples}{
\omega_b&|_{\p\Sigma} = \frac{c}{24\pi}\frac{1}{\rho}\p_\phi(\overline \epsilon \wedge \epsilon)\dr\phi \cr
&+ \frac{c}{48\pi}\left[ 4 L \epsilon' \wedge \epsilon - 4 \overline L \overline \epsilon' \wedge\overline \epsilon + \overline\epsilon' \wedge \epsilon'' + \overline \epsilon'' \wedge \epsilon' - \epsilon'''\wedge\epsilon + \overline\epsilon''' \wedge \overline\epsilon \right]\dr\phi + \mathcal{O}(\rho).
}
Since we will integrate this over the circle to form the symplectic form we are free to drop total derivatives. In particular this means the divergent term above may be dropped. Furthermore, in the $\mathcal{O}(\rho^0)$ term the two terms which mix $\epsilon$ and $\overline\epsilon$ combine to a total derivative. So the symplectic form factorizes into a holomorphic and antiholomorphic term as expected,
\eq{24 examples}{
\Omega = \frac{c}{48\pi}\int_0^{2\pi}\left[ (4 L \epsilon' - \epsilon''') \wedge \epsilon - (4 \overline L \overline\epsilon' - \overline\epsilon''') \wedge \overline \epsilon \right]\dr\phi.
}

But recalling \eqref{4 examples} and \eqref{6 examples} we have
\eq{25 examples}{
\delta L \wedge \epsilon = \frac{1}{2}(4 L \epsilon' - \epsilon''') \wedge \epsilon
}
and similarly for $\overline L$. Of course, we identify the RHS here as being precisely the factor which appears in \eqref{24 examples}. Making the replacement we recover the first line of \eqref{12 examples}.

\subsubsection{New AdS boundary conditions}\label{New AdS GR}

Here we again consider the Einstein-Hilbert Lagrangian \eqref{1 examples} in $D = 3$, but this time we consider the modified boundary conditions described in \cite{Compere:2013bya}. Many of the details in the calculations here are similar to those which appear in topologically massive gravity, to be discussed in Section \ref{Application to Topologically Massive Gravity}, so the manipulations here will be useful as a warmup in addition to a demonstration of the techniques in Section \ref{Computing Gauge Orbit Actions}. We write an arbitrary Fefferman-Graham gauge metric as
\eq{26 examples}{
\dr s^2 = \frac{\dr\rho^2}{4\rho^2} + \frac{1}{\rho}\left( g_{ab}^{(0)} + \rho g_{ab}^{(1)} + \mathcal{O}(\rho^2)\right)\dr x^a \dr x^b
}
where again $\rho > 0$ is the radial coordinate with asymptotic boundary at $\rho$ and $x^a = (t, \phi)$ are the coordinate on the boundary with $\phi \sim \phi + 2\pi$. As before we will use the coordinates $w = t + \phi$ and $\overline w = -t + \phi$ for convenience\footnote{These coordinates relate to the coordinates $(r, t^+, t^-)$ which appear in \cite{Compere:2013bya} by $\rho = 1/r^2$ and $t^+ = w$, $t^- = -\overline w$. Later we will meet a pair of functions $(P, L)$ which were $(\overline P, \overline L)$ in \cite{Compere:2013bya}.}.

The boundary conditions considered in \cite{Compere:2013bya} can be phrased as fixing the components
\eq{27 examples}{
g_{\overline w \overline w}^{(0)} = 0,\ \ \ \ \p_{\overline w}g_{ww}^{(0)} = 0,\ \ \ \ g_{w\overline w}^{(0)} = \frac{1}{2},\ \ \ \ g_{\overline w \overline w}^{(1)} = 4 G\Delta
}
with $\Delta$ a fixed constant. Importantly, one can check that, while it is slightly modified from the Gibbons-Hawking-York boundary term, there indeed exists a choice of $\ell$ such that $B = 0$ in \eqref{4 canonical}.

The Einstein equations can be solved exactly for metrics obeying these boundary conditions. The analog of the Ba\~nados metrics for this case are
\eq{28 examples}{
\dr s^2 =& \frac{\dr \rho^2}{4 \rho^2} + \frac{1}{\rho}\dr w (\dr \overline w + \p_w P\dr w) + 4G [ L \dr w^2 + \Delta (\dr \overline w + \p_w P \dr w)^2] \cr
&+ (4G)^2 \Delta \rho L \dr w(\dr \overline w + \p_w P \dr w)
}
where $P = P(w)$ and $L = L(w)$ are free functions parametrizing the solutions.

Furthermore, the asymptotic vector fields which preserve these boundary conditions are given by
\eq{29 examples}{
\xi^\rho &= \rho \epsilon'\cr
\xi^w &= \epsilon + \mathcal{O}(\rho^2) \cr
\xi^{\overline w} &= \sigma - \frac{1}{2}\rho \epsilon'' + \mathcal{O}(\rho^2)
}
where $\epsilon(w)$ and $\sigma(w)$ are free functions parametrizing the diffeomorphism.

These diffeomorphisms act on the parameters $P$ and $L$ by
\eq{30 examples}{
i_{V_\xi}\delta P &= \sigma + P' \epsilon, \cr
i_{V_\xi}\delta L &= 2 L \epsilon' + \epsilon L' - \frac{1}{8G}\epsilon'''.
}
The charges associated to these diffeomorphisms can be computed to be
\eq{33-1 examples}{
H[\epsilon, \sigma] = \int_0^{2\pi}\frac{\dr\phi}{2\pi} \left[ \left(\Delta P'^2 - L\right) \epsilon + \Delta(1 + 2 P') \sigma \right].
}

From \eqref{30 examples} it's clear that the $\epsilon$ part of the transformation acts just like \eqref{6 examples} on $L$ while the $\sigma$ part of the transformations acts to shift $P$. This means we can integrate up \eqref{30 examples} to find
\eq{31 examples}{
P = g(w),\ \ \ \ L = - \frac{1}{8G}\{f, w\}.
}
This can be done explicitly by applying the finite diffeomorphism
\eq{32 examples}{
\rho' &= f' \rho + G\Delta \frac{f''^2}{f'}\rho^3 + \mathcal{O}(\rho^4), \cr
w' &= f + 2 G \Delta f'' \rho^2 + \mathcal{O}(\rho^3), \cr
\overline w' &= \overline w + g - \frac{1}{2}\frac{f''}{f'}\rho + \mathcal{O}(\rho^3),
}
where $f(w)$ and $g(w)$ are free functions, to the background metric, which we choose to be an extremal BTZ black hole,
\eq{33 examples}{
\dr s^2 = \frac{\dr\rho'^2}{4 \rho'^2} + \frac{1}{\rho'}\dr w' \dr \overline w' + 4G\Delta \dr \overline w^2.
}

Diffeomorphisms act on the orbit parameters by
\eq{34 examples}{
i_{V_\xi} \delta f = f' \epsilon, \ \ \ \ i_{V_\xi}\delta g = \sigma + g' \epsilon,
}
which can be found by composing diffeomorphisms or deducing the transformation rule from \eqref{30 examples} and \eqref{31 examples}. Using \eqref{31 examples} we may also write the momentum as
\eq{33-1-1 examples}{
P &= H[1, 1] = \Delta + \int_0^{2\pi}\frac{\dr\phi}{2\pi}\left[ \Delta ( 2 + g') g' + \frac{1}{8G}\{f, w\} \right] \cr
&= \Delta + \int_0^{2\pi}\frac{\dr\phi}{2\pi}\left[ \Delta  g' g' -\frac{1}{2} \frac{1}{8G}\left(\frac{1}{f'}\right)'' f' \right]
}
where in the second line we have used that $\{f, w\} = -\frac{1}{2}\left(\frac{1}{f'}\right)''f'$ up to the addition of total derivatives. As was the case in \eqref{9-1-1 examples}, this momentum is of the form \eqref{eeb} for the momentum method. The canonical 1-form is given by
\eq{33-2 examples}{
\Theta = \int_0^{2\pi}\frac{\dr\phi}{2\pi}\left[ \Delta g' \delta g - \frac{1}{16G}\left(\frac{1}{f'}\right)'' \delta f \right].
}
From this and the Hamiltonian $H_t = H[1, -1]$ we obtain the phase space action
\eq{33-3 examples}{
S = \int\frac{\dr t\dr\phi}{2\pi}\left[ \Delta g' (\dot g - g') - \frac{1}{16 G}\left(\frac{1}{f'}\right)''(\dot f - f') \right].
}

To instead apply our other techniques, \eqref{34 examples} supplies the analog of \eqref{11 examples}, now taking the form
\eq{35 examples}{
\left(\begin{array}{c}
\delta f\\
\delta g
\end{array}\right) = \left(\begin{array}{cc}
f' & 0\\
g' & 1
\end{array}\right)\left(\begin{array}{c}
\epsilon_\delta\\
\sigma_\delta
\end{array}\right).
}

With all of these expressions for this theory we are well positioned to compute the boundary symplectic form via the methods in Sections \ref{iWTheta} and \ref{Noether Charge}. To make things even easier, we may note that \eqref{22 examples} applies for any theory governed by the Einstein-Hilbert Lagrangian, independent of the boundary conditions we impose, so long as $B = 0$, which happens to be the case for this theory.

Performing the evaluation we find that again the divergent $1/\rho$ term is a total derivative as it was for asymptotically Dirichlet boundary conditions while the finite term can be massaged to yield
\eq{36 examples}{
\Omega = \int_0^{2\pi}\frac{\dr\phi}{4\pi}\left[ \delta(L - \Delta P'^2) \wedge \epsilon_\delta - 2\Delta \delta P' \wedge \sigma_\delta \right],
}
which one may check has canonical 1-form \eqref{33-2 examples}. Furthermore, comparing this expression to the the charges \eqref{33-1 examples} it's immediately clear that the prescription \eqref{1-5 ivomega} reproduces \eqref{36 examples} as well. Thus both the techniques of Sections \ref{iWTheta} and \ref{Noether Charge} lead to the same phase space action \eqref{33-3 examples}.

\subsection{Comparison of techniques}

We have presented several techniques for computing actions governing boundary modes.  It is useful to discuss the computational advantages offered by each approach.

When applicable, the momentum method is the most efficient approach.  As presented, it applies to $D=3$ diffeomorphism orbits built on bulk solutions that are translationally invariant in the boundary directions. The needed inputs are simply expressions for the boundary Hamiltonian and momentum charges written in terms of the orbit variables.  The prescription involves using integration by parts to write the momentum in a canonical form, after which the canonical 1-form $\Theta$ can be read off.  Together with the expression for the Hamiltonian, the boundary action follows.    A great advantage here is that one gets $\Theta$ directly, bypassing the need to solve $\delta \Theta=\Omega$.  Also noteworthy is that it can be applied order by order in perturbation theory, for example by expanding in powers of the boundary fluctuations.   This method proved very effective in \cite{Ebert:2022cle}.

The other methods we discussed are less efficient but have a wider range of applicability. Using the transfer vector to invert $i_V\Omega = -\delta H_V$ requires that we we have computed all of the Noether charges, rather than just the momentum. Of course this has the advantage that it is applicable in any dimension and with any type of gauge orbit. Additionally, while the computation of the Noether charges will generally be sensitive to the boundary conditions of the theory, once the charges are computed the actual computation \eqref{1-5 ivomega} is immediate. As seen in the examples \eqref{12 examples} and \eqref{36 examples}, the resulting form for $\Omega$ may not make solving $\delta\Theta = \Omega$ immediate.

The transfer field method is likely the most computationally involved of the three approaches, but has the unique advantage of offering a clean separation between contributions depending on the Lagrangian of the theory and the boundary conditions imposed. In the case one wishes to study a Lagrangian under a variety of boundary conditions, this method could become more efficient than the others which would require that we recompute the Noether charges with each set of boundary conditions. For example, once \eqref{22 examples} was known for $D = 3$ Einstein-Hilbert theory, there was essentially no additional computation required to apply it in the example in Section \ref{New AdS GR} beyond working out the new class of asymptotic vector fields.

As a final point here, we note that while computing the Noether charges generally involves solving a condition \eqref{14 canonical} to obtain the charges, each step in the transfer field method can in principle be completed algorithmically. This opens the possibility that, given $\theta$ and $k_w$ for the theory, the computation of both $Q_w$ and $\tilde k_w$ can be automated on the computer, leaving only the computation of $B$ to obtain an expression for \eqref{12 1formV}.

\section{Application to Topologically Massive Gravity}\label{Application to Topologically Massive Gravity}

Topologically massive gravity \cite{Deser:1981wh} is described by the Einstein-Hilbert action supplemented with a gravitational CS term,
\eq{wza}{S_{TMG}  =  \frac{1}{16\pi G} S_{EH}  - \frac{l }{96\pi G \nu} S_{CS} }
with\footnote{It is straightforward to check that the boundary action may be chosen such that $B = 0$ in \eqref{4 canonical}.}
\eq{wzb}{ S_{EH} &= \int\! \dr^3x \sqrt{g} \left(R+\frac{2}{l^2}\right) +S_{bndy}  \cr
S_{CS} & = \int \Tr \left( \Gamma\wedge  d\Gamma +\frac{2}{3} \Gamma \wedge \Gamma \wedge \Gamma \right)~.}
Here the connection 1-form is $\Gamma^\alpha_{~\beta}  =\Gamma^\alpha_{\beta \mu} \dr x^\mu$, where $\Gamma^\alpha_{\beta \mu}$ are the usual Christoffel symbols built out of the metric.  We take $\nu>1$ and henceforth set $l=1$.  Our interest in applying our general techniques to this theory is to illustrate various issues that are not present in simpler examples, in particular the the non-diffeomorphism invariance of the action and the existence of solutions with relatively exotic ``warped" asymptotics.  We find that these pose no obstacle to implementing our general procedure to find the boundary action.

We will be particularly interested in solutions \cite{Anninos:2008fx} that are the warped analog of the more familiar Ba\~nados geometries \cite{Banados:1998gg},
\eq{wzc}{\dr s^2 = L^2 \Bigg[\frac{\dr r^2}{r^2}  +u^2 \left( \dr t+\Kh \dr \phi+(r+r^{-1}\Lch)\dr \phi\right)^2-(r-r^{-1} \Lch)^2 \dr \phi^2    \Bigg]~,\quad \phi \cong \phi+2\pi  }
where
\eq{wzd}{L^2 = \frac{1}{\nu^2+3}~,\quad u^2 =\frac{4\nu^2 }{\nu^2+3}~.}
\rf{wzc} is a solution to $S_{TMG}$ for any functions $\Kh=\Kh(\phi)$ and  $\Lch=\Lch(\phi)$.  Setting $\Kh=\Lch=0$ gives the analog of Poincar\'e AdS$_3$; constant values of $(\Kh,\Lch)$ can be obtained by quotienting the vacuum solution (analogous to how one obtains BTZ); $(\Kh,\Lch)$ with nontrivial dependence on $\phi$ (analogous to the Ba\~nados geometries)  are obtained by applying asymptotic symmetry transformations to the solutions with constant values.

The solutions \rf{wzc} live in a phase space defined by boundary conditions that are preserved by the asymptotic coordinate transformations $x^\mu \rt x^\mu +\chi^\mu$ which  have the large $r$ behavior
\eq{wzh}{ \chi^\phi&= \xi^\phi +\frac{1}{ 2r^2}\p_\phi^2 \xi^\phi+ \ldots  \cr
\chi^t & = \xi^t -\frac{1}{r} \p_\phi^2 \xi^\phi + \ldots\cr
\chi^r & = -r \p_\phi \xi^\phi + \ldots}
Here $\xi^{t,\phi}=\xi^{t,\phi}(\phi)$ are arbitrary periodic functions of $\phi$.  Under an infinitesimal diffeomorphism of the form  \rf{wzh} we have
\eq{wzt}{  \delta \Lch & =  -\frac{1}{ 2} \p_\phi^3 \xi^\phi +2 \Lch \p_\phi \xi^\phi + \p_\phi \Lch \xi^\phi \cr
\delta \Kh & = \p_\phi \xi^t + \p_\phi (\Kh \xi^\phi) }
Associated to any boundary preserving vector field $\xi = \xi^\mu \p_\mu$ is a charge  $H[\xi]$.   For example, the charges associated with rigid  time and angular translations, i.e energy and angular momentum, are  
\eq{wzg}{ M & =\frac{1}{uL}H[ \p_t]=  \frac{1}{ 12\pi G}\int_0^{2\pi} \! \Kh \dr\phi   \cr
J & =H[\p_\phi]=  -\frac{1}{ 6\pi G }  u L\int_0^{2\pi} \!  \left[ \left( 1+ \frac{1}{ u^2}  \right)\Lch -\frac{1}{ 4}  \Kh^2   \right] \dr\phi }
as will be derived below.\footnote{Original references for the derivation of the charges and asymptotic symmetry algebras include \cite{Nazaroglu:2011zi,Bouchareb:2007yx,Compere:2008cv,Compere:2009zj,Blagojevic:2009ek}.}

To derive the charges we can consider $S_{EH}$ and $S_{CS}$ independently of each other.   That is, the action $S= C_1 S_{EH} +C_2 S_{CS}$ yields charges $Q'=C_1 Q'_{EH} + C_2 Q'_{CS}$, so we can extract $Q'_{EH}$ and $Q'_{CS}$ by setting one of $C_{1,2}=0$.   This is allowed because the procedure to obtain $Q$ is linear, and the current $J_{EH}$ is conserved in the $S_{EH}$ theory, and likewise for $J_{CS}$.

\subsection{Einstein-Hilbert contribution}    

As in \eqref{14 examples} we have
\eq{wzga}{ Q^{EH}_\chi = -\varepsilon^{\alpha \beta}_{~~\phi} \nabla_{\alpha}\chi_\beta \dr\phi}
Evaluating this on \rf{wzc} gives
\eq{wzgb}{ Q^{EH}_\chi & = Lu(u^2-1) r^2 \xi^\phi +2Lu(u^2-1)  \Kh r \xi^\phi +Lu^3r \xi^t  \cr
& \quad +Lu \Big[u^2(2\Lch +\Kh^2) +6\Lch \Big]\xi^\phi +Lu^3 \Kh \xi^t  + O(r^{-1})  }
The term proportional to $r^2$ is constant on the phase space and so will be omitted in what follows.  The same goes for the last term on the first line.   On   the other hand, there is a linearly diverging term in the first line which is not constant on phase space; this term will cancel a similar term in the CS contribution, yielding a finite result as $r \rt \infty$.

\subsection{CS contribution}

The Lagrangian 3-form $L_{CS} = \Tr (\Gamma \wedge \dr\Gamma +\frac{2}{3} \Gamma^3)$  has variation 
\eq{wzhz}{ \delta L_{CS}  =   E^{\gamma \rho}  \delta g_{\gamma\rho}\sqrt{-g} \dr^3x + \dr\theta^{CS} }
with
\eq{wzi}{ E^{\gamma \rho} & = -  \nabla_\beta R^{\beta\rho}_{~~\mu\nu} \varepsilon^{\gamma\mu\nu}  = -2 C^{\gamma\rho}  \cr
\theta^{CS} &= \Tr( \delta \Gamma \wedge \Gamma)  -  2\delta g_{\kappa\rho} R^\rho_{~\delta} \dr x^\kappa \wedge \dr x^\delta~,}
where $C^{\gamma\rho}$ is the Cotton tensor and the Ricci tensor $R_{\mu\nu}$ is formed from the Riemann tensor $R= \dr \Gamma + \Gamma^2$ as $R_{\mu\nu} = R^\alpha_{~\mu\alpha\nu}$.  We also recall that in $D=3$ the Riemann tensor can be expressed in terms of the Ricci tensor; in terms of the Riemann two-form this amounts to the identity
\eq{wzia}{R^\alpha_{~\beta} & = \left[ \delta^\alpha_{\gamma} R_{\delta\beta}-g_{\beta\gamma}R^\alpha_{~\delta}- \frac{1}{ 2}R \delta^\alpha_{\gamma} g_{\delta\beta} \right]  \dr x^\gamma \wedge \dr x^\delta   }

Following the discussion in appendix \ref{Charge Calculation} to compute the charges, the Noether current corresponding to the bulk vector field $\xi$ is
\eq{wzj}{ J^{CS}_\xi =i_{V_\xi}\theta^{CS} -i_\xi L_{CS} - Y_\xi~. }
where $Y_\xi$ defined via
\eq{wzk}{ \dr Y_\xi &= \delta_\xi L_{CS}-\Lc_\xi L_{CS} \cr
&= \Tr (\dr v \wedge \dr \Gamma) }
with $v^\alpha_{~\beta} = \p_\beta \xi^\alpha$.   We take
\eq{wzl}{ Y_\xi =-\Tr(\dr v \wedge \Gamma)~.}
We also have
\eq{wzm}{ i_{V_\xi}\theta^{CS} &= \nabla_\beta \nabla_\gamma \xi^\alpha dx^\beta \wedge \Gamma^\gamma_{~\alpha} + \Tr ( i_\xi R \wedge \Gamma)  -2 (\nabla_\kappa \xi_\rho+\nabla_\rho \xi_\kappa) R^\rho_\delta \dr x^\kappa \wedge \dr x^\delta
 \cr
i_\xi L_{CS} & = \Tr ( i_\xi \Gamma \dr\Gamma- \Gamma \wedge i_\xi R )}
Using these relations along with \rf{wzia}, some algebra leads to
\eq{wzn}{ J^{CS}_\xi = dQ^{CS}_\xi}
with
\eq{wzo}{ Q^{CS}_\xi
& =  \p_\gamma \xi^\alpha \Gamma^\gamma_{~\alpha}  + \nabla_\gamma \xi^{\alpha} \Gamma^\gamma_{~\alpha}  +  \xi^a \big[  - 4 R_{a \delta} +R g_{a \delta} \big]\dr x^\delta~.  }
The full charge is 
\eq{wzp}{ H_\xi^{CS} = \int_{\p \Sigma} \left[ Q^{CS}_\xi - C^{CS}_\xi \right] }
where $C^{CS}_\xi$ is obtained by solving $\delta C^{CS}_\xi = i_\xi \theta$, which can be written explicitly as 
\eq{wzja}{ \delta C^{CS}_\xi = (\delta \Gamma^\alpha_{\kappa \beta}\Gamma^\beta_{\delta \alpha} - \delta \Gamma^\alpha_{\delta \beta}\Gamma^\beta_{\kappa \alpha}) \xi^\kappa \dr x^\delta  -2(\delta g_{\kappa \rho}  R^\rho_{~\delta} - \delta g_{\delta \rho} R^\rho_{~\kappa} ) \xi^\kappa \dr x^\delta  }
To proceed further we need to specify the asymptotic form of the solutions of interest.   Taking the solutions to be of the form \rf{wzc} it is straightforward to compute
\eq{wzq}{  Q^{CS}_\xi & =\Big\{ Lu(u^2-1) r^2 \xi^\phi+2 u(u^2-1)L \Kh r \xi^\phi   +u(u^2-1) L r \xi^t\cr
&  + L \Big[  u\left(u^2-\frac{2}{ 3}\right) \Kh^2 +\left(2u^3-\frac{10}{ 3} u +\frac{8}{ 3u} \right)\Lch \Big]\xi^\phi + Lu \left( u^2-\frac{2}{ 3}\right) \Kh \xi^t   + O(r^{-1}) \Big\} \dr\phi \cr
  C^{CS}_\xi &  =\Big\{ -u\left(u^2-\frac{2}{ 3}\right)L \Kh \xi^t + O(r^{-1})\Big\} \dr\phi
 }
Again, we can ignore terms that are constant on phase space.

\subsection{Total charge}

The total charge is   
\eq{wzr}{ H_\xi =\frac{1}{16\pi G} \int_{\p \Sigma} \left[ Q_\xi -\frac{1}{6 \nu} (Q^{CS}_\xi- C^{CS}_\xi)\right]~.
}
Noting the cancellation between (non-constant) diverging term, we arrive at a finite result in the large $r$ limit,
\eq{wzs}{ H_\xi =  \frac{1}{16\pi G} \int_{\p \Sigma}\left[ \frac{4}{ 3}uL \Kh \xi^t- \frac{8}{ 3} u L \left( \big( 1+\frac{1}{ u^2}  \big)\Lch -\frac{1}{ 4}  \Kh^2   \right) \xi^\phi\right] \dr\phi~.}

\subsection{Lower spin gravity formulation}

While Einstein gravity in three dimensions can be recast as a CS theory (with gauge group $SL(2,\RR)\times SL(2,\RR) $), this no longer holds in the presence of a gravitational CS term because the theory is no longer topological.  However, the subsector of the theory described by the solutions \rf{wzc}  can be described by a CS theory, namely one with gauge group $SL(2,\RR)\times U(1)$, so-called ``lower spin gravity" \cite{Hofman:2014loa,Azeyanagi:2018har}.  The choice of gauge group is dictated by the isometry group of the warped vacuum solution.   In particular, there is a fairly simple relation such that the charges and symplectic form of the two theories are mapped to each other.  Since the equations of CS theory are much easier to deal with than those of TMG, this provides a simpler route to isolating the boundary degrees of freedom.   The simplification arises essentially because the bulk degrees of freedom have been omitted. 

 The action is  
 \eq{wzu}{ S= \frac{ k}{ 4\pi} \int \Tr \left(A\wedge\dr A +\frac{2}{ 3} A\wedge A\wedge A\right)+\frac{\kb}{ 8\pi} \int \Ab \wedge \dr\Ab }
 Here $A$ is an $SL(2,\RR)$  connection and $\Ab$ is a $U(1)$ connection.   The $SL(2,\RR)$ generators obey $[L_m,L_n]=(m-n)L_{m+n}$ and we use the two-dimensional representation, for which $\Tr L_1 L_{-1}=-1$. 
 \eq{wzv}{ a &= (L_1 -\Lch L_{-1})\dr \phi + (\omega_1 L_0 + \omega_2 L_{-1})\dr t \cr
 \overline{a} & = K \dr \phi+ \dr t~.}
 Here, as is standard, the boundary connections $(a,\overline{a})$ are related to bulk connections $(A,\Ab)$ by a gauge transformation that encodes the radial dependence, 
 \eq{wzva}{A=b^{-1}ab+b^{-1}\dr b~,\quad \Ab= b\overline{a}b^{-1}+b\dr b^{-1}~,}
 where $b=e^{-\frac{1}{2}L_0 \ln r }$.    Gauge transformations act as
 \eq{wzw}{ \delta a = \dr\eps +[a,\eps]~,\quad \delta \overline{a}= \dr\overline{\eps} }
 with $\eps = \eps_1 L_1 + \eps_0 L_0 +\eps_{-1}L_{-1}$.  The form of \rf{wzv} is preserved by taking $\eps_0=-\p_\phi \eps_1$ and $\eps_{-1}= \frac{1}{ 2} \p_\phi^2 \eps_1 -\Lch \eps_1$,   To connect to the metric description we trade $(\eps_1,\overline{\eps})$ for $(\xi^\phi,\xi^t)$ according to 
 \eq{wzx}{\eps_1 = \xi^\phi~,\quad  \overline{\eps}= \xi^t  + \Kh \xi^\phi~,}
 with
 \eq{wzxa}{ \Kh = \frac{4\pi}{\kb}K~.}
 The gauge transformations then act as
\eq{wzy}{  \delta \Lch & =  -\frac{1}{ 2} \p_\phi^3 \xi^\phi +2 \Lch \p_\phi \xi^\phi + \p_\phi \Lch \xi^\phi \cr
\delta \Kh & = \p_\phi \xi^t + \p_\phi (\Kh \xi^\phi) }
reproducing \rf{wzt}.

We can now apply relations reviewed in Appendix \rff{CSapp} for CS theory.  Letting $V$ denote the phase space vector field implementing an infinitesimal  gauge transformation with  parameters $(\eps,\overline{\eps})$ we have
$i_V \Omega = -\delta H[\eps,\overline{\eps}]$ with
\eq{hyz}{  H[\eps,\overline{\eps}] &= \frac{k}{2\pi}\int_{\p \Sigma}  \Tr( \eps a) + \frac{\kb}{4\pi}\int_{\p \Sigma}  \overline{\eps} \ab \cr
& =  \int_0^{2\pi} \! \dr\phi \left[ \left(\frac{k}{2\pi}\Lch+\frac{\kb }{8\pi}\Kh^2\right) \xi^\phi + \frac{\kb }{4\pi}\Kh \xi^t \right] \cr
& = \int_0^{2\pi} \! \dr\phi (  \Lc \xi^\phi + K\xi^t  )~,  }
where we have defined $\Lc$ according to
\eq{hyza}{ \Lch = \frac{2\pi}{ k} (\Lc - \frac{2\pi}{ \kb} K^2)~.}
The second line of \rf{hyz} agrees with \rf{wzs} under the identification
\eq{hyzab}{ k=  \frac{uL}{3G}\left(1+\frac{1}{u^2}\right)~,\quad \kb=-\frac{uL}{3G}~.}
We conclude that there is a simple relation between the canonical structure of lower spin gravity and the class of solutions \rf{wzc}  to TMG.   

One can also build  the metrics \rf{wzc}  out of the $SL(2,\RR)\times U(1) $ connection, in analogy with the corresponding relation in ordinary 3D gravity.   See \cite{Hofman:2014loa} for details.    Gauge transformations in the CS theory will map to diffs in the metric formulation, as follows from the fact that we have already mapped the transformations on the phase space variables $(\Lch,\Kh)$.   

\subsection{Boundary action for warped \texorpdfstring{AdS$_3$}{AdS3}} 

To compute the boundary action our first task is to obtain expressions for the charges evaluated on a given orbit.  The finite diff functions are written as $(\Phi,T)$ with
\eq{caa}{ \Phi = \phi+\xi^\phi~,\quad T=t+\xi^t~.}
The infinitesimal transformations are given in \rf{wzy}.  The first expression is familiar and exponentiates via the Schwarzian derivative and the second expression is easily handled to yield
\eq{cab}{ \Lch(\phi) & = (\p_\phi \Phi)^2 \Lch_0-\frac{1}{2} \{\Phi(\phi),\phi\} \cr
\Kh(\phi) & = \p_\phi \Phi \Kh_0 +\p_\phi T(\Phi) }
where the constant values $(\Lch_0,\Kh_0)$ serve as parameters labelling the orbit under consideration.   Translating to unhatted variables, as appear in the charge expression $Q= \int\! \dr\phi (\Lc \xi^\phi+ K \xi^t)$ we have 
\eq{cac}{ \Lc(\phi)& =  (\p_\phi \Phi)^2 \Lc_0 -\frac{k}{ 4\pi } \{\Phi,\phi\} +\p_\phi \Phi \p_\phi T K_0+\frac{\overline{k}}{ 8\pi} (\p_\phi T)^2   \cr
K(\phi) &=  \p_\phi \Phi  K_0 + \frac{\overline{k}}{ 4\pi} \p_\phi T~.   }

 To extract the boundary symplectic potential $\Theta $ we use the momentum method.  The momentum, i.e. generator of $\phi$ translations, is 
 \eq{cad}{ P = H[\xi= \p_\phi] =  \int_0^{2\pi} \Lc(\phi) \dr\phi~. }
To apply the momentum method  we are instructed to write $P$ in the form 
\eq{cae}{ P = \int(P_\Phi \Phi' + P_T T')\dr\phi}
which is achieved by taking\footnote{We should note that $\Lc(\phi)$ has several terms which contain only first derivatives; as discussed in section \rf{momsec}, since they are purely quadratic their contribution to $\Omega$ is easily found by directly solving $i_V\Omega = -\delta Q_V$ for such terms.  } 
\eq{caf}{  P_\Phi &= \Lc_0 \Phi' +\frac{k}{ 8\pi} \left(\frac{1}{ \Phi'}\right)''\cr
 P_T &  = K_0 \Phi' +\frac{\overline{k}}{ 8\pi} T'~.  }
The symplectic potential is then 
 \eq{cag}{ \Theta = \int \! \dr\phi \left[  \left(  \Lc_0 \Phi' +\frac{k}{ 8\pi} \left(\frac{1}{ \Phi'}\right)'' \right) \delta \Phi + \left( K_0 \Phi' +\frac{\overline{k}}{ 8\pi} T' \right) \delta T   \right]~.  }
The Hamiltonian corresponding to time translations is
\eq{cah}{ H = Q[\xi=\p_t] = \int K \dr\phi =  2\pi K_0~, }
so the kinetic term in the action is obtained by making the replacements $(\delta \Phi ,\delta T ) \rt (\dot{\Phi},\dot{T})$.    
The fact that this  $H$ is a constant on phase space reflects the fact the solutions in this theory are time independent.    More interesting dynamics are obtained by choosing a Hamiltonian that generates translations along the vector field $ \p_t + \Omega \p_\phi$.   This is appropriate for computing a partition function of the form $\Tr e^{-\beta (H + \Omega P)}$.   Such partition functions are the appropriate ones to consider for two reasons \cite{Detournay:2012pc}.  First, unitary representations of the warped current  algebra  have $H$ unbounded from below while $P$ is positive definite, so $\Omega >0$ is required for convergence.  Second, the black hole solutions \rf{wzc}  (with constant $(\Lch,\Kh)$) have horizon Killing vector field  (defined to be the vector field that vanishes at the bifurcation surface) 
\eq{cai}{ \xi_H &= \p_t + \Omega \p_\phi \cr
& = \p_t  + \frac{1}{2\sqrt{\Lch}+ \Kh} \p_\phi }
$\Omega$ must be positive for a smooth horizon, as can be seen from the expression for the surface gravity 
\eq{caj}{ \kappa = \frac{\sqrt{\Lch}}{\sqrt{\Lch}+\Kh/2}~.  }

As in all of our examples, the boundary action captures all of the information regarding the warped spacetimes that is dictated by symmetry.  For instance, it can be used to compute correlators of the currents, as well one loop corrections to the partition function coming from the boundary modes.

\section{Emergent Boundary Modes}
\label{emergent}

As developed so far, boundary  field theories arise due to the imposition of boundary conditions that are preserved by some infinite  dimensional group of transformations that act nontrivially at the boundary.   However, situations can arise in which the boundary in question is not a sharp boundary in the sense of terminating the space on which the theory lives, but rather an interface that connects the original region to an ``outer region".    The question is whether or not the boundary modes survive; that is, are they effectively transported to the outer region?

This situation arises naturally in the AdS/CFT context.   For example, one might have an AdS$_3$ solution which appears as the near horizon limit of some non-asymptotically AdS$_3$ solution $M$.  Is the asymptotic symmetry group of the AdS$_3$ region visible at the boundary of the larger spacetime $M$?   From the dual CFT point of view, the AdS$_3$ region encodes physics in the deep IR, so one is asking about how to these IR degrees of freedom are realized in terms of the UV description.

We will flesh out how this works in a particular example.  Five-dimensional Einstein-Maxwell theory (with $\Lambda<0$) admits an asymptotically AdS$_5$ solution that has a near horizon AdS$_3 \times \RR^2$ factor which we  compactify to  $T^2$ \cite{DHoker:2009mmn}.  The $T^2$ is supported by a nonzero Maxwell field strength.     We show how the boundary graviton phase space is visible at the AdS$_5$ boundary.  Including also a CS term $A\wedge F \wedge F$, the AdS$_3$ region supports  boundary photons, and we again show how these appear in the full description.    These results are consistent with the low energy correlators computed in    \cite{DHoker:2010xwl}, which are sensitive to the emergent boundary modes.  For related work see \cite{Adawi:1998ta,Cederwall:1998tr}.

\subsection{Background solution}

We first briefly summarize the relevant features of the solution; more details may be found in \cite{DHoker:2009mmn}.   The action is
\eq{jaa}{ S=-\frac{1}{ 16\pi G_5} \int\! \dr^5x \sqrt{g} \left(R +F^{MN}F_{MN} -12\right) +\frac{k}{ 12\pi G_5} \int A\wedge F \wedge F  }
We consider a solution of the form
\eq{jac}{ \dr s^2 &= \frac{\dr r^2}{ L(r)^2} +2L(r) \dr x^+ \dr x^- +e^{2V(r)} \dr x^i \dr x^i~,\quad i=1,2\cr
F &=b \dr x^1 \wedge \dr x^2  }
The function $L$ can be found in terms of $V$, but numerics are required to find $V$. We will only need the asymptotics.  As $r \rt \infty$ we have AdS$_5$ asymptotics,
\eq{jae}{ L(r)&= 2r +\ldots~,\quad 
e^{2V(r)}   =  c_V r+ \ldots  }
for some constant $c_V$. 

As $ r \rt 0$ we have AdS$_3\times T^2$ asymptotics, 
\eq{jaf}{ L(r)  = 2br + \ldots~,\quad 
e^{2V(r)} & = 1  +\ldots }
Fluctuations of the metric and gauge field  with polarizations and spacetime dependence restricted to the AdS$_3$ directions are governed by an effective 3d Einstein-CS theory, so if we place a boundary in this region we will find the usual 2d boundary photons and gravitons.   On the other hand, the asymptotic symmetry group at the AdS$_5$ boundary is finite dimensional, which leads to the aforementioned question of how the near horizon boundary modes are visible in terms of observables computed at the AdS$_5$ boundary. 

\subsection{Boundary photons} 

\subsubsection{Linearized solutions} 

We considered a linearized gauge field perturbation of the form 
\eq{jag}{\delta A = a_+(r,x^+) \dr x^+~.}
It obeys
\eq{jah}{ L e^{2V}  \frac{ \p a_+ }{ \p r}+-2kb a_+ =0 }
The solution which is smooth at the origin (assuming $k>0$) is 
\eq{jai}{ a_+(r,x^+) = \eps(x^+)e^{-2kb \int_r^\infty  \frac{\dr r'}{L(r')e^{2V(r')}} }  }
This has asymptotics:
\eq{jam}{  & r \rt 0:\quad\quad a_+(r,x^+) = \eps(x^+) r^k + \ldots \cr
 & r \rt \infty :\quad\quad a_+(r,x^+) =  \eps(x^+) \left( 1-\frac{kb}{c_V r}+ \ldots  \right)~.   }
 The  $1/r$ falloff term implies a nonzero boundary current $J_+ \sim \eps(x^+)$. 
 This solution is not quite what we want however, since normalizable solutions, corresponding to vanishing source in the dual CFT, should vanish in the large $r$ limit.  We can remedy this by performing a gauge transformation.   We first of all write 
 \eq{jama}{ \eps(x^+)= -\p_+ \lambda(x^+) }
 and then perform a gauge transformation with parameter $\Lambda(r,x^+) = f(r) \lambda(x^+)$ where 
\eq{japc}{\lim_{r \rt 0} f(r)=0~,\quad \lim_{r\rt \infty} f(r) =1~.}
 After the gauge transformation we have
\eq{japd}{a_++\p_+\Lambda  & = \p_+ \lambda(x^+)\left(f(r)-e^{-2kb \int_r^\infty  \frac{dr'}{L(r')e^{2V(r')}} }  \right) \cr
a_r+\p_r\Lambda  & = \lambda(x^+) \p_r f(r)\cr
a_-+\p_-\Lambda  & = 0 }
such that all components  vanish at both small and large $r$, as desired. It is apparent that these modes are not pure gauge, as they carry a nonzero field strength.

\subsubsection{Symplectic form}

We now wish to compute the symplectic form restricted to the space of these linearized solutions.  In particular, we compute the full $D=4+1$ symplectic form for these non-pure gauge solutions.  We will see that the result agrees with what we would get by considering pure gauge modes living in the near horizon AdS$_3$, with a boundary imposed in that region.  

Since all polarizations and spacetime dependence of the fluctuations is confined to three dimensions it is convenient to dimensionally reduce the action \rf{jaa}. Keeping the gauge field terms, after integrating over the $T^2$ the action can be written as 
\eq{jape}{S= \frac{bV_2}{4\pi G_5} \int_{M_3}  L }
where $V_2$ is the coordinate volume of $T^2$ and 
\eq{japf}{ L = \frac{1}{2}  \Phi F\wedge \star F +k A\wedge F }
where $\Phi=b^{-1} e^{2V}$.  We will consider $S=\int L$ and tack on the $\frac{bV_2}{4\pi G_5}$  prefactor at the end.   Proceeding as usual, we vary the action with respect to the gauge field and write the result as in \rf{2 canonical}, $\delta L =  E_A \delta A + \dr \theta$ yielding the field equation
\eq{japg}{\dr [ \Phi \star \dr A +2kA]=0  }
and the symplectic form $\Omega = \int_\Sigma \omega $  with
\eq{jbe}{ \omega = \delta \theta =  -\Phi \delta A \wedge \star \dr \delta A -k\delta A \wedge \delta A }
The fluctuation \rf{japd}  takes the form $A= a+\dr \Lambda$ where $a$ obeys
\eq{jbh}{ \Phi \star \dr a +2ka =0~.}
Using this we have 
\eq{jbk}{ \Omega =  k\int_{\Sigma}  \big(\delta a \wedge \delta a  - \dr \delta \Lambda \wedge \dr \delta \Lambda \big)~. }
In the case of interest $a=a_+ \dr x^+$ and so the first term vanishes, while the second terms is an exact form.  Restoring the prefactor, we then  obtain the symplectic form
\eq{jbm}{ \Omega=  -\frac{kb V_2}{ 4\pi G_5}  \int_{\p \Sigma}  \delta \Lambda \wedge \dr \delta \Lambda~. }
This symplectic form lives on the AdS$_5$ boundary.   However, and this is the main point, the form of the result is precisely the same as would have been obtained by just considering the pure CS action $\frac{kb }{4\pi G_5} \int A \wedge F$ defined on the near horizon AdS$_3$ region.    It is furthermore easy to verify that the boundary currents take the same form as well.    This shows very explicitly how the near horizon boundary photon degrees of freedom are effectively transported to the AdS$_5$ boundary.  The mechanism for this is not entirely trivial; in particular, note that the fluctuation modes we used in this analysis are not pure gauge in the near horizon region, yet the symplectic form defined on them agrees with that of pure gauge modes in AdS$_3$.

\subsection{Boundary gravitons}

\subsubsection{Linearized solutions} 

We now consider a fluctuation around the metric in \rf{jac} by considering 
\eq{jap}{ \dr s^2 & = \frac{\dr r^2}{ L^2} +2L\dr x^+ \dr x^- +M (\dr x^+)^2 +e^{2V} \dr x^i \dr x^i }
where we will work to linear order in $M=M(r,x^+)$.  For $r \ll 1$ background metric contains the AdS$_3$ factor $\dr s_3^2 = \frac{\dr r^2 }{4b^2 r^2}+4br \dr x^+ \dr x^-$.  In this near horizon region, a boundary graviton is obtained by applying a diff $x^\mu \rt x^\mu +\xi^\mu$ with 
\eq{japz}{ \xi = \eps(x^+)\p_+ -\p_+ \eps(x^+) r \p_r - \frac{1}{8b^3 r} \p_+^2 \eps(x^+)\p_-~. }
This yields a fluctuation of the form 
\eq{japy}{ M= -\frac{1}{2b^2} \p_+^3 \eps(x^+)~.}
The fluctuation \rf{japy} is not a solution of the linearized Einstein equations outside the near horizon region.  Our task is to extend \rf{japy} to a solution in the full spacetime that respects the AdS$_5$ asymptotics.  As shown in \cite{DHoker:2010xwl}  such a solution is obtained by multiplying \rf{japy} by $-2bL^c(r)$, where 
\eq{japw}{ L^c(r) = L(r) \int_{\infty}^r \frac{\dr r'}{ L(r')^2 e^{2V(r')} }~.}
The asymptotics of this function are 
\eq{jak}{r\rt 0\quad\quad  L_0^c(r) &  \sim -\frac{1}{ 2b} \cr
r\rt \infty \quad\quad  L_0^c(r) &  \sim -\frac{1}{ 4c_V r}~.}
So the desired fluctuation mode is
\eq{jakz}{ M = b^{-1} L^c(r) \p_+^3 \eps(x^+)~.}
This non-pure gauge solution carries a nonzero AdS$_5$ boundary stress tensor ($c=3/2G_3$ is the near horizon Brown-Henneaux central charge),
\eq{jaky}{ T_{++} = -\frac{c_V c }{96\pi} \p_+^3 \eps(x^+) }
which is the same form as the contribution to the AdS$_3$ boundary stress tensor coming from the pure gauge mode \rf{japy}. 

\subsubsection{Symplectic form} 

The fluctuation modes found above are not yet in a form suitable for computing the symplectic form.  Indeed, contracting two such perturbations with the general gravitational symplectic form will give zero; this follows from symmetry considerations, as each perturbation carries two lower $+$ indices and there are no available upper $+$ indices to soak these up.  The issue is that the perturbation is singular at $r=0$ due to the breakdown of coordinates there.  We can fix this by performing a compensating diff that zeroes out the perturbation as $r\rt 0$ but acts trivially at large $r$.  Let $f(r)$ interpolate smoothly between $-1$ and $0$ in going from small to large $r$.  We then act with a diff of the form \rf{japz} except with $\eps(x^+)$ replaced by $f(r)\eps(x^+)$.  We write the combined perturbation as 
\eq{jakw}{ \phi + \phi_f~.}
By construction, it vanishes as $r\rt 0$ and is equivalent to our original perturbation at large $r$.     The symplectic form contracted against two such perturbations is\footnote{Here we are using the notation $\Omega(\phi_1,\phi_2)=i_{V_2}i_{V_1}\Omega$, where $V_i$ denotes the phase space vector field corresponding to the linearized perturbation $\phi_i$.} 
\eq{jakx}{ \Omega(\phi_1+\phi_{1f}, \phi_2+\phi_{2f})~.}
 Now, $\Omega(\phi_1,\phi_2)=0$ as already noted.   Also $\Omega(\phi_{1f},\phi_2+\phi_{2f})=0$.   This follows since $\phi_{1f}$ is a pure diff mode, and $\Omega$ contracted against a pure diff mode localizes to the boundary of $\Sigma$, but $\phi_2+\phi_{2f}$ vanishes as $r\rt 0$ and $\phi_{1f}$ vanishes as $r\rt \infty$, so both boundary terms are zero.

All that survives is therefore $\Omega(\phi_{1},\phi_{2f})$.  This again localizes to the boundary, but now we get a nonzero boundary term at small $r$,   Since $\phi_1 =-\phi_{1f}$ at small $r$, this   boundary term is precisely the same expression as appears in the symplectic form of pure diff modes in AdS$_3$.\footnote{This relative minus sign is cancelled by the minus sign that occurs when we switch from taking the boundary to be inner versus outer.   } So we once again find that the perturbations defined on the full spacetime carry the same symplectic form as the pure gauge near horizon modes.

\section*{Acknowledgements}
We thank Ruben Monten for useful discussions.
P.K. is supported in part by the National Science Foundation grant PHY-2209700.

\appendix

\section{Forms Conventions}

For convenience we collect some conventions and useful expressions involving forms which are used elsewhere in this work. For a general $p$-form,
\eq{1 form}{
\omega = \frac{1}{p!}\omega_{\mu_1\cdots\mu_p}\dr x^{\mu_1}\wedge\cdots\wedge \dr x^{\mu_p}, \ \ \ \ \dr\omega &= \frac{1}{p!}\p_\nu \omega_{\mu_1\cdots\mu_p}\dr x^\nu \wedge \dr x^{\mu_1} \wedge \cdots \wedge \dr x^{\mu_p} \cr
&= \frac{1}{p!}\nabla_\nu\omega_{\mu_1\cdots\mu_p}\dr x^\nu \wedge \dr x^{\mu_1}\wedge\cdots\dr x^{\mu_p}
}
so long as the connection $\nabla$ is torsionless. This also implies
\eq{2 form}{
\frac{1}{p!}\p_{[\nu}\omega_{\mu_1\cdots\mu_p]} = \frac{1}{(p+1)!}(\dr\omega)_{\nu\mu_1\cdots\mu_p}
}
where the antisymmetrization is defined to include a division by the order of the symmetric group\footnote{For example these conventions imply $(\dr\omega)_{\alpha\beta} = \p_\alpha\omega_\beta - \p_\beta\omega_\alpha$.}. The contraction is defined by
\eq{3 form}{
i_\xi \omega = \frac{1}{(p-1)!}\xi^\nu\omega_{\nu\mu_2\cdots\mu_p}\dr x^{\mu_2}\wedge\cdots\wedge \dr x^{\mu_p}, \ \ \ \ (i_\xi\omega)_{\mu_2\cdots\mu_p} = \xi^\nu\omega_{\nu\mu_2\cdots\mu_p}.
}

We often find it useful to define
\eq{5 form}{
(\dr^{D-p}x)_{\mu_1\cdots\mu_p} &= \frac{1}{(D-p)!}\epsilon_{\mu_1\cdots\mu_p\nu_1\cdots\nu_{D-p}}\dr x^{\nu_1}\wedge\cdots\wedge\dr x^{\nu_{D-p}}
}
and it is useful to note
\eq{5-1 form}{
i_\xi(\dr^{D - p}x)_{\mu_1\cdots\mu_p} &= \xi^\nu(\dr^{D-(p+1)}x)_{\mu_1\cdots\mu_p\nu}.
}
With these notations the Hodge star is defined by
\eq{4 form}{
\star\omega = \frac{\sqrt{|g|}}{p!(D - p)!}\omega^{\mu_1\cdots\mu_p}\epsilon_{\mu_1\cdots\mu_p\nu_{p+1}\cdots\nu_D}\dr x^{\nu_{p+1}}\wedge\cdots\wedge\dr x^{\nu_D} = \frac{\sqrt{|g|}}{p!}\omega_{\mu_1\cdots\mu_p}(\dr^{D-p}x)^{\mu_1\cdots\mu_p}
}
where $\epsilon_{\mu_1\cdots\mu_D}$ is the totally antisymmetric numeric array with entries $\pm1$. With this definition $\star^2 = \sgn(g)(-1)^{p(D - p)}$.

One may also show, assuming torsionless connection, that for $p + 1 \not= D$
\eq{6 form}{
\nabla_\nu X^{\mu_1\cdots\mu_{D-(p+1)}\nu}(\dr^{p+1}x)_{\mu_1\cdots\mu_{D-(p+1)}} = \dr\left( \frac{1}{D - (p+1)}X^{\mu_1\cdots\mu_{D-(p+1)}\nu}(\dr^p x)_{\mu_1\cdots\mu_{D - (p+1)}\nu} \right).
}
Note there is no factorial in the coefficient. A useful special case of this is $p = D - 2$ (so $\dr\omega$ is a $(D-1)$-form current) is
\eq{7 form}{
\nabla_\nu X^{\mu\nu}(\dr^{D-1}x)_\nu = \dr\left(\frac{1}{2}X^{\mu\nu}(\dr^{D - 2}x)_{\mu\nu} \right).
}
For the special case $p = D$, if $\omega = \nabla_\mu j^\mu \sqrt{|g|}\dr^Dx$, then $\omega = \dr J$ with $J = \star j = \sqrt{|g|}j^\mu(\dr^{D-1}x)_\mu$.

Finally, suppose $\Sigma$ is a codimension-1 (non-null) surface in $M$ and let $\phi^*$ be the pullback to $\Sigma$, as would appear in an integral over $\Sigma$. Then
\eq{8 form}{
\phi^*\left[ \sqrt{|g|} (\dr^{D-1}x)_\mu \right] = \sigma \hat n_\mu \Vol(\Sigma)
}
where $\hat n_\mu$ is the unit normal to $\Sigma$ and $\sigma = \hat n^\mu \hat n_\mu$ depends on whether $\Sigma$ is timelike or lightlike. The orientation choice here is $\Vol(M) = \hat n \wedge \Vol(\Sigma)$.

\section{Identically Closed Forms}\label{Ap identically closed forms}
\label{ident}

The Poincar\'e lemma ensures that if $J$ is any closed form on spacetime then, at least locally, there exists a potential $Q$ satisfying $J = \dr Q$. Though such conserved currents are common in physics due to Noether's theorem, the Poincar\'e lemma is not typically useful because it carries no guarantee that $Q$ can be constructed locally from the fields in our theory. Indeed, for a general current $J$, $Q$ will indeed be non-local. The exception to this, of course, are the Noether currents associated to gauge symmetries. These are currents $J_\xi$ which are closed for every possible set of free functions $\xi$ parametrizing the gauge transformation. This additional ingredient, closure despite depending on an arbitrary function, is what we need to ensure the potential $Q_\xi$ can be constructed as a local functional of $\xi$ and the other fields in the theory, assuming $J_\xi$ was local to begin with.

This result has appeared in the physics literature in \cite{Wald:1990} and \cite{Barnich:2000zw, Barnich:2001jy}, the latter referring to the older mathematical literature on the bivariational complex\footnote{See \cite{Olver:1986, Anderson:1989} for textbook treatments on this perspective.} where this result is established by finding a homotopy operator. While \cite{Wald:1990} is clear on the recursive algorithm for constructing the potential $Q_\xi$, insights from the more mathematical literature on the utility of higher Euler operators allow us solve the recursion explicitly in a natural way\footnote{This is essentially deriving the homotopy operator, though we will only show that it constructs local potentials and not that it obeys the stronger property of defining a homotopy operator.}.

As most of this work only relies on the existence of an algorithm to compute a local $Q_\xi$ and not on its details, we state the result as a theorem here\footnote{The original statement in \cite{Wald:1990} is slightly more general as the $\xi$ are allowed to be sections of an arbitrary bundle rather than functions, but this is all we will require and allows us to include the statement about $\xi = 0$.}:
\newtheorem{theorem}{Theorem}
\begin{theorem}\label{Wald Theorem}
As elsewhere, suppose $M$ is a $D$-dimensional spacetime, $\phi$ is some collection of fields over $M$, and $\xi$ are some functions over $M$. Let $J$ be a $p$-form over $M$ ($p<D$) and a local functional of $\phi$ and $\xi$, meaning at each point $x\in M$ it depends on only $\phi$ and $\xi$, and finitely many of their derivatives, at $x$. Suppose further that $\dr J = 0$ for all free functions $\xi$ and that $J = J_0$ when $\xi = 0$. Then there exists a local functional $Q$ of $\phi$ and $\xi$ such that $J = J_0 + \dr Q$.
\end{theorem}

For the interested reader, we have included a short but pedagogical review of how the algorithm can be derived, by way of a simple example, in section \ref{Motivation}. The statement of the algorithmic procedure can be found in section \ref{idclosed General Algorithm}, along with a small notational dictionary to some other locations in the literature.

\subsection{Motivation}\label{Motivation}

The idea underlying this algorithm, in this presentation, relies on a simple observation which can be demonstrated by considering the special case where $J$ is a 2-form which depends only linearly on the functions $\xi$ up to the first derivative. It is also conceptually simpler to start with a potential for $J$, say $\tilde Q$, rather than $J$ itself\footnote{This $\tilde Q$ will be related to the potential $Q$ we construct in the end, but in general will be different. The actual algorithm makes no reference to this $\tilde Q$, so it's only purpose here is as a useful intermediary to motivate the key ideas.}. Writing $\tilde Q$ as an expansion in the derivative order of $\xi$ we have
\eq{1 idclosed}{
\tilde Q &= \left[\tilde Q_{k\alpha}\xi^k + \tilde Q^\mu_{k\alpha}\p_\mu \xi^k\right]\dr x^\alpha \cr
&= \left[ \xi^k E_k(\tilde Q_{\alpha}) + \p_\mu[\xi^k E^\mu_k(\tilde Q_{\alpha})] \right] \dr x^\alpha
}
where in the second line we have integrated by parts until no derivatives act directly on the $\xi^k$. We will refer to the coefficients $E_k^{\mu_1\cdots\mu_r}(\tilde Q_\alpha)$ as the Euler coefficients\footnote{This language is inherited from the mathematical literature where the $E_k^{\mu_1\cdots\mu_r}$ would be called the higher Euler operators as they are generalizations of the Euler-Lagrange operator which produces the equations of motion from a Lagrangian.}. Here we would have
\eq{2 idclosed}{
E_k(\tilde Q_{\alpha}) = \tilde Q_{k\alpha} - \p_\mu \tilde Q^\mu_{k\alpha},\ \ \ \ E_k^\mu(\tilde Q_{\alpha}) = \tilde Q^\mu_{k\alpha}.
}
Since the $\xi^k$ are free functions, the coefficients in the derivative expansion are unique. Since there are only finitely many derivatives present we can always convert between the derivative and Euler expansions by integration by parts identities, so the Euler expansion must also be unique\footnote{This is the step which requires the ``finite derivative order" part of locality. In \cite{Wald:1990} it was that the recursion started with the highest derivative order.}.

Of course, this is merely a rewriting of $\tilde Q$, but it's a rewriting with the advantage that when we take the differential of $\tilde Q$ we find
\eq{3 idclosed}{
\dr \tilde Q &= \frac{1}{1!}\big[&&\ \ \ \ \ \ \ \ \ \ \ \ \  0 &+&& \p_\rho[\xi^k E_k(\tilde Q_\alpha)] & &+& &\p_\rho\p_\mu[\xi^k E_k^\mu(\tilde Q_\alpha)] &\big]\dr x^\rho \wedge \dr x^\alpha \cr
&= \frac{1}{2!}\big[& &\xi^k E_k(\dr \tilde Q_{\rho\alpha}) &+&& \p_\mu[\xi^k E_k^\mu(\dr \tilde Q_{\rho\alpha})] & &+& & \p_\mu\p_\nu[\xi^k E^{\mu\nu}_k(\dr \tilde Q_{\rho\alpha})] &\big] \dr x^\rho \wedge \dr x^\alpha
}
where in the second line we have written the generic Euler expansion for $\dr \tilde Q = J$. But because the Euler expansion is unique, the aligned terms must match so
\eq{4 idclosed}{
\frac{1}{2!}E_k(J_{\rho\alpha}) = 0,\ \ \ \ \frac{1}{2!}E_k^\mu(J_{\rho\alpha}) = \delta^\mu_{[\rho}E_k(\tilde Q_{\alpha]}),\ \ \ \ \frac{1}{2!}E_k^{\mu\nu}(J_{\rho\alpha}) = \delta^{(\mu}_{[\rho}E^{\nu)}_k(\tilde Q_{\alpha]}).
}
Throughout this section, antisymmetrization will not include the $k$ index as it's generically of a different type. If we were able to invert these equations for the Euler coefficients of $\tilde Q$, we would have determined $\tilde Q$ in terms of $J$.

Unfortunately, we cannot invert these equations, but we don't need to because $J$ doesn't have a unique potential. So really since our goal is to construct a potential for $J$, we only need to reconstruct $\tilde Q$ up to a closed form. To do this we want to eliminate the Kronecker deltas, and we may do so by taking a trace between $\mu$ and $\rho$ in \eqref{4 idclosed}. We find
\eq{5 idclosed}{
\frac{1}{2!}E^\mu_k(J_{\mu\alpha}) &= \frac{1}{2!}\left(D E_k(\tilde Q_\alpha) - E_k(\tilde Q_\alpha)\right), \cr
\frac{1}{2!}E^{\mu\nu}_k(J_{\mu\alpha}) &= \frac{1}{2!2!}\left( D E_k^\nu(\tilde Q_\alpha) - E_k^\nu(\tilde Q_\alpha) + E_k^\nu(\tilde Q_\alpha) - \delta^\nu_\alpha E_k^\mu(\tilde Q_\mu) \right).
}
The overall factors come from the (anti-)symmetrization and we see that we can classify the types of terms which appear. If either $\mu$ or $\rho$ (or both) appear on the delta, we obtain a term proportional to an Euler coefficient of $\tilde Q$. If both indices do not appear on the delta, then we obtain a term which still contains a delta. In the case both $\mu$ and $\rho$ appear on the delta we find the coefficient to be the dimension. If only $\mu$ is on the delta we obtain a minus sign\footnote{That this observation generalizes to higher forms and more derivatives of $\xi$ is not completely obvious, but is nonetheless true. The proof is essentially an exercise in combinatorics.\label{combinatorics comment}}, if only $\rho$ is on the delta then we obtain a plus sign.

We may rewrite \eqref{5 idclosed} in the form
\eq{6 idclosed}{
E_k(\tilde Q_\alpha) &= \frac{1}{D - 1}E^\mu_k(J_{\mu\alpha}). \cr
E_k^\nu(\tilde Q_\alpha) &= \frac{2}{D}E_k^{\mu\nu}(J_{\mu\alpha}) + \frac{1}{D} \delta_\alpha^\nu E^\mu_k(\tilde Q_\mu).
}
The second equation does not specify the Euler coefficient\footnote{In fact, the inability to invert these equations is related to the fact that we are really defining a homotopy operator. If the homotopy operator is denoted $h$, then $\tilde Q = h\dr \tilde Q + \dr h \tilde Q = h J + \dr h \tilde Q$. The ``delta terms'' in our organization of the contraction collects into the second term here as we argue shortly. See e.g. \cite{Olver:1986} for a proof that these terms organize into the particular form $\dr h \tilde Q$. These extra terms are not important to our particular application.} of $\tilde Q$ in terms of only $J$, but observe that if we try to build $\tilde Q$ back from these expressions we find
\eq{7 idclosed}{
\tilde Q = \left[ \frac{1}{D - 1}\xi^k E^\mu(J_{\mu\alpha}) + \p_\nu\left(\xi^k \left[ \frac{2}{D}E^{\mu\nu}_k(J_{\mu\alpha}) + 2 \delta^\nu_\alpha E^\mu_k(\tilde Q_\mu)\right] \right) \right] \dr x^\alpha.
}
So the term preventing us from completely determining the Euler coefficients of $\tilde Q$ in terms of those of $J$ has collected itself into an exact form! Furthermore, we can see that this will be a generic conclusion because all of the upper spacetime indices (besides $\mu$ which is already contracted) have to contract on derivatives when we construct the form from the Euler coefficients. In the ``delta type'' terms, at least one of these derivatives will therefore contract on the delta and hence on a form index, making the term a total differential.

This means we can always write $\tilde Q = Q + \dr \alpha$ where $\alpha$ is constructed from all of the ``delta type'' terms. Since $Q$, which is thus constructed entirely from the Euler coefficients of $J$, is related to $\tilde Q$ by an exact form, both are equally good potentials for $J$ and so there is no loss in taking $Q$ over $\tilde Q$.

The generalization to higher form degrees and arbitrary derivative orders in $\xi$ is now mostly a matter of making sure we get the coefficients on the various terms in the contraction correct. This is not completely trivial, but a straightforward combinatorics problem, the main difficulty resting with the result mentioned in Footnote \ref{combinatorics comment}.

\subsection{General algorithm}\label{idclosed General Algorithm}

Suppose that $J$ is an identically closed $p$-form ($p < D$) depending on a set of free functions $\xi^k$ and finitely many derivatives thereof. We give the algorithm in three steps. First we state it for the special case where $J$ depends only linearly on $\xi$ and its derivatives. We then point out the simple generalization to other derivative operators. Finally, we state how the algorithm for the linear case also gives the solution to the non-linear case.

Since $J$ is linear in $\xi$ and its derivatives, it can always be written in the form
\eq{8 idclosed}{
J &= \left[ J_{k \alpha_1\cdots\alpha_p}\xi^k + J_{k\alpha_1\cdots\alpha_p}^\mu \p_\mu \xi^k + \cdots \right]\dr x^{\alpha_1}\wedge \cdots\wedge \dr x^{\alpha_p} \cr
&= \left[ \xi^k E_k(J_{\alpha_1 \cdots \alpha_p}) + \p_\mu[\xi^k E^\mu_k(J_{\alpha_1\cdots\alpha_p})] + \cdots \right]\dr x^{\alpha_1}\wedge \cdots\wedge \dr x^{\alpha_p}
}
where in the second line we have integrated by parts to remove all derivatives from $\xi$. The Euler coefficients, $E^{\mu_1\cdots\mu_r}_k(J_{\alpha_1 \cdots\mu_p})$, are all the data we need to construct a local potential for $J$. Note that the upper indices of the Euler coefficients are assumed to be symmetrized.

With this, the potential is given by
\eq{9 idclosed}{
Q = \frac{1}{(p - 1)!}\sum_{r = 0}\frac{r + 1}{D - p + r + 1}\p_{\mu_1}\cdots\p_{\mu_r}[\xi^k E_k^{\nu\mu_1\cdots\mu_r}(J_{\nu\alpha_2\cdots\alpha_p})] \dr x^{\alpha_2}\wedge\cdots\dr x^{\alpha_p}
}
where the upper bound on the sum over $r$ depends on the derivative order $J$ as it's a sum over all of $J$'s Euler coefficients. We also take the convention where $r = 0$ indicates no derivatives should be taken. The factor of $(p - 1)!$ can be absorbed into a contraction against the vector $\p_\nu$ to write
\eq{10 idclosed}{
Q = \sum_{r = 0} \frac{r + 1}{D - p + r + 1}\p_{\mu_1}\cdots\p_{\mu_r}[\xi^k E^{\nu\mu_1\cdots\mu_r}_k(i_{\p_\nu}J)].
}

This formula appears elsewhere in the literature using slightly different notation. In \cite{Barnich:2001jy, Compere:2007az} a multi-index notation is used (multi-index notation is also employed in \cite{Olver:1986, Anderson:1989}, but the conventions are slightly different which makes coefficients look different) wherein $(\mu)$ is a tuple of indices, $|\mu|$ is its length, and $((\mu)\nu)$ is the concatenation of the tuple $(\mu)$ with the additional index $\nu$. If we further denote $E^{\mu_1\cdots\mu_r}_k = \frac{\delta}{\delta \xi^k_{(\mu)}}$ and $i_{\p_\nu}J = \frac{\p J}{\p\dr x^\nu}$ then \eqref{10 idclosed} may be written compactly as
\eq{11 idclosed}{
Q = \frac{|\mu| + 1}{D - p + |\mu| + 1}\p_{(\mu)}\left[ \xi^k \frac{\delta}{\delta \xi^k_{(\nu(\mu))}}\frac{\p J}{\p\dr x^\nu} \right]
}
where a summation over the length of the tuple $(\mu)$ is to be understood. For example, this is equation (A36) in \cite{Compere:2007az}.

The generalization from partial derivatives to covariant derivatives of arbitrary type is described in \cite{Wald:1990}. In the proof of \eqref{9 idclosed} the only property of the derivatives used is that they obey the product rule and that the upper indices of the Euler coefficients are completely symmetric. Symmetry of the indices follows trivially for partial derivatives, but for covariant derivatives we can always work with the symmetrized derivatives at the cost of introducing field strengths. So if we assume that all derivatives have first been symmetrized, we can write\footnote{Though we use the same symbol for the Euler coefficient, the Euler coefficients computed using symmetrized covariant derivatives and using partial derivatives will, of course, not generally be the same.}
\eq{12 idclosed}{
J = \left[ \xi^k E_k(J_{\alpha_1 \cdots \alpha_p}) + \nabla_\mu[\xi^k E^\mu_k(J_{\alpha_1\cdots\alpha_p})] + \cdots \right]\dr x^{\alpha_1}\wedge \cdots\wedge \dr x^{\alpha_p}
}
with potential
\eq{13 idclosed}{
Q = \frac{1}{(p - 1)!}\sum_{r = 0}\frac{r + 1}{D - p + r + 1}\nabla_{\mu_1}\cdots\nabla_{\mu_r}[\xi^k E_k^{(\nu\mu_1\cdots\mu_r)}(J_{\nu\alpha_2\cdots\alpha_p})]\dr x^{\alpha_2} \wedge \cdots \wedge \dr x^{\alpha_p}.
}

For the generalization to non-linear dependence of $J$ on $\xi$ we consider an arbitrary 1-parameter path $\xi(\lambda)$ through $\xi$-space. Denote by $J_\lambda$ the evaluation of $J$ on $\xi(\lambda)$. This $J_\lambda$ is still identically closed and so it's $\lambda$ derivative is as well:
\eq{14 idclosed}{
\dr \dot J_\lambda = 0
}
where we have denoted the $\lambda$ derivative by a dot.

Now $\dot J_\lambda$ is an identically closed form which depends only linearly on $\dot \xi$, which is arbitrary and independent of $\xi(\lambda)$. Hence we may apply the \eqref{13 idclosed} for the linear case to $\dot J$ using $\dot \xi$ as our free function to find
\eq{15 idclosed}{
Q_\lambda = \sum_{r = 0}\frac{r + 1}{D - p + r + 1}\nabla_{\mu_1}\cdots\nabla_{\mu_r}[\dot \xi^k E_k^{(\nu\mu_1\cdots\mu_r)}(i_{\p_\nu}J)]
}
which satisfies $\dot J_\lambda = \dr Q_\lambda$. The Euler coefficients now, of course, must be understood to have been computed from integrating by parts on $\dot\xi^k$.

Taking the $\lambda$ integral we find
\eq{16 idclosed}{
J_1 = J_0 + \dr \int_0^1Q_\lambda\dr\lambda.
}
If we choose the flow $\xi(\lambda)$ such that $\xi(0) = 0$ and $\xi = \xi(1)$, then theorem \ref{Wald Theorem} is an immediate consequence.

A useful choice of path is the contracting flow\footnote{Much like the Poincar\'e contracting homotopy, this choice is not invariant under redefinitions of the $\xi$. Furthermore, there may be cases where this path is not suitable. We refer the reader to the discussion in \cite{Wald:1990} for some considerations in this direction. The contracting flow is always possible when the $\xi$ are functions on spacetime.} $\xi(\lambda) = \lambda \xi$. In this case the potential for $J$ is
\eq{17 idclosed}{
Q = \int_0^1\dr\lambda \sum_{r = 0}\frac{r + 1}{D - p + r + 1}\nabla_{\mu_1}\cdots\nabla_{\mu_r}[\xi^k E^{(\nu\mu_1\cdots\mu_r)}_k(i_{\p_\nu}J(\lambda\xi))]
}
where it should be understood that the Euler coefficients used here should be those of $\dot J_\lambda$. Making the notational changes described above \eqref{11 idclosed}, we see that this expression is (A.9) in \cite{Barnich:2001jy}.

\subsection{Examples}

As a simple first example we may derive the Komar term \eqref{14 examples}. Using \eqref{1 examples} and \eqref{2 examples} we find
\eq{18 idclosed}{
J_\xi = \frac{\sqrt{-g}}{16\pi G}\left( \nabla_\nu(\nabla^\mu \xi^\nu + \nabla^\nu \xi^\mu) - 2\nabla^\mu\nabla_\nu \xi^\nu \right)(\dr^{d}x)_\mu - i_\xi L.
}
The potential is simple to compute in this case without using the heavy machinery introduced above. We need only write
\eq{19 idclosed}{
\nabla^\mu\nabla_\nu \xi^\nu = -R^\mu_\nu\xi^\nu + \nabla_\nu \nabla^\mu \xi^\nu
}
so
\eq{20 idclosed}{
J_\xi = \left(\frac{\sqrt{-g}}{8\pi G}R^\mu_\nu \xi^\nu(\dr^d x)_\mu - i_\xi L\right) + \frac{\sqrt{-g}}{16\pi G}\nabla_\nu(\nabla^\nu \xi^\mu - \nabla^\mu \xi^\nu)(\dr^dx)_\mu.
}
The first pair of terms, linear in $\xi$ with no derivatives thereof, vanish on the equations of motion. The latter pair of terms can be identified as yielding the Komar term via \eqref{7 form}.

To demonstrate the heavy machinery, note first that we only need the linear case \eqref{10 idclosed} and that only Euler derivatives with at least one upper index (meaning at least one derivative in the Euler expansion of $J_\xi$) contribute. So any terms linear in $\xi$ with no derivatives, after putting $J_\xi$ in Euler form, will not contribute. Indeed, we can see in the example \eqref{4 idclosed} that these terms must vanish when $J_\xi$ is closed\footnote{This is actually a familiar statement. Note that the zeroth Euler coefficient is precisely the Euler-Lagrange operator acting on $J_\xi$. This is then the standard statement that the Euler-Lagrange equations annihilate total derivatives.}. In particular, this means the $i_\xi L$ term will not contribute to $Q_\xi$, which we also found in \eqref{20 examples}.

Since only the $i_{V_\xi}\theta$ terms in $J_\xi$ will contribute. In the remaining terms we need to symmetrize the covariant derivatives so the Euler coefficients will have symmetrized upper derivatives. Here the derivatives are all 2nd order, so the anti-symmetrization will produce Riemann curvatures with no additional derivatives on $\xi$. These terms will again not contribute. The result is now
\eq{21 idclosed}{
J_\xi = \frac{\sqrt{-g}}{16\pi G}\nabla_{(\alpha}\nabla_{\beta)}\left[ \left( g^{\beta\mu}\delta^\alpha_k + g^{\alpha\beta}\delta^\mu_k - 2g^{\alpha\mu}\delta^\beta_k \right) \xi^k \right](\dr^dx)_\mu + (\cdots)_k \xi^k.
}
The relevant Euler coefficients are then
\eq{22 idclosed}{
E^\alpha_k(J_\xi) = 0,\ \ \ \ E^{\alpha\beta}_k(J_\xi) = \frac{1}{2}\left( 2 g^{\alpha\beta}\delta^\mu_k - g^{\alpha\mu}\delta^\beta_k - g^{\beta\mu}\delta^\alpha_k \right)(\dr^{d}x)_\mu.
}

The reconstruction \eqref{10 idclosed} now produces
\eq{23 idclosed}{
Q_\xi &= \frac{\sqrt{-g}}{16\pi G}\left\{ 0 + \frac{1 + 1}{D - (D - 1) + 1 + 1}\nabla_\alpha\left[ \xi^k E^{\alpha\beta}_k(i_{\p_\beta}J_\xi) \right] \right\} \cr
&= \frac{\sqrt{-g}}{16\pi G}\frac{2}{3}\frac{1}{2}\nabla_\alpha\left[ \xi^k \left( g^{\alpha\beta}\delta^\mu_k - g^{\alpha\mu}\delta^\beta_k - g^{\beta\mu}\delta^\alpha_k \right) (\dr^{d-1}x)_{\mu\beta} \right]
}
which is again the Komar term, though this algorithm is obviously not unnecessary for this example.

Another example, this time non-linear in $\xi$, would be the calculation \eqref{15 examples}. This example is again simple enough that we can simply obtain the result \eqref{16 examples} without applying the general algorithm. To do this we need only \eqref{8 1formV} to find
\eq{24 idclosed}{
16\pi G\delta k_\xi &= \frac{4\Lambda}{D - 2}\sqrt{-g}\left[ \xi^\nu \wedge \nabla_\nu \xi^\mu + \nabla_\nu\xi^\nu \wedge \xi^\mu \right] (\dr^{d}x)_\mu
}
which is \eqref{16 examples} upon collecting the total derivative and using \eqref{7 form}.

To find this result using the general algorithm, we first take a derivative of $\delta k_\xi$ to obtain
\eq{25 idclosed}{
16\pi G\delta\dot k_\xi &= \frac{4\Lambda}{D - 2}\sqrt{-g}\nabla_\nu\left( \dot \xi^\nu \wedge \xi^\mu + \xi^\nu \wedge \dot\xi^\mu \right) (\dr^{d}x)_\mu \cr
&= \frac{4\Lambda}{D - 2} \sqrt{-g}\nabla_\nu\left[ \dot\xi^k\wedge(\delta^\nu_k \xi^\mu - \delta^\mu_k\xi^\nu) \right] (\dr^dx)_\mu.
}
From this we extract the Euler coefficient
\eq{26 idclosed}{
E^\nu_k(\delta \dot k_\xi) = \frac{1}{16\pi G}\frac{4\Lambda}{D - 2}\sqrt{-g}\wedge \left( \delta^\nu_k\xi^\mu - \delta^\mu_k\xi^\nu \right)(\dr^dx)_\mu
}
where we have included a dangling wedge to remind ourselves that when we perform the potential reconstruction the $\dot\xi$ appears to the left of this coefficient.

Since $\delta k_\xi = 0$ when $\xi = 0$, we have \eqref{16 idclosed} with $J_0 = 0$ if we use the contracting flow \eqref{17 idclosed}. The reconstruction \eqref{17 idclosed} then yields
\eq{27 idclosed}{
16\pi G\tilde k_\xi &= \int_0^1\dr\lambda\frac{0 + 1}{D - (D - 1) + 0 + 1}\frac{4\Lambda}{D - 2}\sqrt{-g}\xi^k\wedge \lambda(\delta^\nu_k\xi^\mu - \delta^\mu_k\xi^\nu)(\dr^{d -1}x)_{\mu\nu} \cr
&= -\frac{2\Lambda}{D - 2}\sqrt{-g}\xi^\mu \wedge \xi^\nu (\dr^{d -1}x)_{\mu\nu},
}
as we have already shown should be the case.

\section{Diffeomorphism Charges for Generally non-Covariant Lagrangians}\label{Charge Calculation}

In section \ref{Symmetries and Charges}, we reviewed a general procedure for computing the Noether charge associated to an arbitrary gauge symmetry. In the special case where the gauge transformation is a diffeomorphism, additional simplifications are possible \cite{Tachikawa:2006sz}.

The transformation of the Lagrangian under a diffeomorphism $\xi$ can generally be separated into covariant and non-covariant pieces as
\eq{1 diff}{
k_\xi = i_\xi L + Y_\xi.
}
We may do the same for the action of the diffeomorphism on $\theta$:
\eq{2 diff}{
\eL_{V_\xi}\theta = \eL_\xi\theta + \tilde\Pi_\xi.
}

With these separations together, \eqref{12 canonical} becomes
\eq{3 diff}{
\dr\Pi_\xi = \dr i_\xi\theta + (\tilde \Pi_\xi - \delta Y_\xi).
}
That is, the covariant part of the transformations collect automatically into a total derivative, so the task of computing $\Pi_\xi$ is reduced to just finding a potential $\Sigma_\xi$ satisfying
\eq{4 diff}{
\dr\Sigma_\xi = \tilde\Pi_\xi - \delta Y_\xi.
}
The computation of $C_\xi$ in \eqref{14 canonical} is now
\eq{5 diff}{
\delta C_\xi = i_\xi \theta + \Sigma_\xi - i_{V_\xi}\delta B.
}

These considerations only alter the computations we need to find $C_\xi$ and do not change the identification of the charge in \eqref{15 canonical} as \eqref{16 canonical}.

\section{Non-Abelian CS}\label{Non-Abelian CS}
\label{CSapp}

Here we consider a non-Abelian CS theory in $D = 3$. This is an important example and stress test of our techniques because of the CS/WZW correspondence which, at first glance, might seem to be in tension with our methods. In particular, recall the well known fact that one must supply an explicit parametrization in order to reduce a WZW model to an explicit boundary theory, and in particular the theory cannot be reduced explicitly to the boundary in terms of the gauge element $g$.

On the other hand, the charges for CS theory can be explicitly constructed on the boundary from $g$ and its derivatives. The construction \eqref{1-5 ivomega} would then seem to suggest that $\Omega$ can also be constructed explicitly from $g$, and we will see that this is indeed the case. In order for our construction \eqref{1 where} of the boundary action to be consistent with this known property of WZW models, it must be the case that while $\Omega$ can be explicitly constructed from $g$, the canonical 1-form $\Theta$ cannot. This is a non-trivial requirement and we will see that it is indeed the case.

So then to define the theory we take the background manifold to be $M = \mathbb{R}\times D$ where the spatial slices are disks. The Lagrangian and some useful quantities are
\eq{a-13 1formV}{
L = \tr\left( A \wedge \dr A + \frac{2}{3}A^3\right),\ \ \ \ E = 2F,\ \ \ \ \theta = -\tr(A \wedge \delta A),\ \ \ \ k_\lambda = \tr(\lambda \dr A).
}
We choose the chiral boundary conditions $A_t = A_\phi$ so there is no need for a boundary contribution to the action, $B = \ell = 0$, in light of \eqref{4 canonical}.

The Noether current associated to gauge transformations is
\eq{a-14 1formV}{
J_\lambda = \tr\left[ \dr(\lambda A) - 2\lambda F\right] = \dr\tr(\lambda A)
}
so $Q_\lambda = \tr(\lambda A)$. Thus one may check that the full Noether charges are
\eq{a-14-1 1formV}{
H[\lambda] = 2\int_0^{2\pi}\dr\phi \tr(\lambda A), \ \ \ \ H_t = \int_0^{2\pi}\dr\phi\tr ( A_\phi^2 ).
}

Before attempting to apply our techniques to this theory, we should use the WZW correspondence to see that the boundary action produced by standard techniques should be. Since we will need an explicit parametrization, and hence gauge group, we wil choose for simplicity $SU(2)$ with the parametrization
\eq{a-14-2 1formV}{
g = e^{i\alpha_1\sigma_1}e^{i\alpha_2\sigma_2}e^{i\alpha_3\sigma_3},\ \ \ \ z = 2\sin(2\alpha_2)
}
where $\sigma_k$ are the Pauli matrices and $z$ turns out to be a convenient definition.

By solving the Lagrange multiplier constraint, the action produced by the Lagrangian \eqref{a-13 1formV} becomes the WZW model
\eq{a-14-3 1formV}{
S = \int_M\tr\left[ \left(g^{-1}\dr g\right)^3 \right] - \int\tr\left[ g^{-1}\p_t g g^{-1}\p_\phi g \right]\dr t\wedge\dr \phi + \int_{\p M}\tr\left[ A_\phi^2 \right]\dr t \wedge \dr \phi.
}
Using $\tr(g^{-1}\dr g)^3 = \dr(z \dr\alpha_1 \wedge \dr \alpha_3)$ the boundary action may be written
\eq{a-14-4 1formV}{
S = -2\int_{\p M}\dr t\dr\phi
\left[(\alpha_1' - z \alpha_3')(\dot\alpha_1 - \alpha_1') + \alpha_2' (\dot\alpha_2 - \alpha_2') + \alpha_3'(\dot\alpha_3 - \alpha_3')\right]
}
where the minus sign comes from identifying the boundary orientation to be $\dr\phi\wedge\dr t$, as commented on in footnote \ref{orientation footnote}.

With this result in mind, since we already have the full Noether charges in hand, the simplest approach is to compute the symplectic form via \eqref{1-5 ivomega}. This produces
\eq{a-14-5 1formV}{
\Omega = \int_0^{2\pi}\tr\left[\delta A_\phi \wedge g^{-1}\delta g \right]\dr\phi = \int_0^{2\pi}\tr\left[ \delta(g^{-1}\p_\phi g) \wedge g^{-1}\delta g \right]\dr\phi.
}
As promised at the beginning of this section, due to the $g^{-1}$ factor, this cannot be written as $\delta$ of something directly in terms of $g$. Instead we must go to the explicit parametrization \eqref{a-14-2 1formV} in order to find the canonical 1-form.

But this is a straightforward calculation to perform, the result being
\eq{a-14-6 1formV}{
\Omega = -2\delta\int_0^{2\pi}\left[ \alpha_1' \delta\alpha_1 + \alpha_2' \delta\alpha_2 + (\alpha_3' - z \alpha_1')\delta\alpha_3 \right] \dr\phi.
}
Together with
\eq{a-14-7 1formV}{
H_t = -2\int_0^{2\pi}\left[ \alpha_1'^2 + \alpha_2'^2 + (\alpha_3' - z\alpha_1')\alpha_3' \right] \dr\phi
}
we find the phase space action
\eq{a-14-8 1formV}{
S = -2\int\dr t\int_{\p\Sigma}\dr\phi\left[\alpha_1' (\dot\alpha_1 - \alpha_1') + \alpha_2' (\dot\alpha_2 - \alpha_2') + (\alpha_3' - z\alpha_1')(\dot\alpha_3 - \alpha_3')\right].
}

We could also find this result by the methods in section \ref{iWTheta}. Since we already have $Q_\lambda$ from \eqref{a-14 1formV}, we only need to compute the potential for $\delta k_w$ with $w = g^{-1}\delta g$. In the case of diffeomorphisms it's simple to use \eqref{6 1formV} as our free function with respect to which the potential $\tilde k_w$ can be constructed. Here, however, $w = g^{-1}\delta g$ is not a completely free function and we must use $\delta\alpha$ as our free function instead. This unfortunately means that we cannot write down a potential $\tilde k_w$ without specifying the gauge group.

Using the parametrization \eqref{a-14-2 1formV} we find
\eq{a-14-9 1formV}{
\tilde k_w = \delta z \wedge \delta \alpha_3 \dr \alpha_1 + \delta\alpha_3 \wedge \delta\alpha_1 \dr z + \delta \alpha_1 \wedge \delta z \dr \alpha_3.
}
Furthermore we may compute
\eq{a-14-10 1formV}{
\delta Q_w =&\ \delta z \wedge (\dr\alpha_3 \delta\alpha_1 + \dr \alpha_1 \delta\alpha_3) + z(\dr\delta\alpha_3\wedge\delta\alpha_1 + \dr\delta\alpha_1 \wedge \delta \alpha_3) \cr
&+ 2(\delta \alpha_1 \wedge\dr\delta\alpha_1 + \delta\alpha_2 \wedge \dr \delta\alpha_2 + \delta\alpha_3 \wedge \dr \delta\alpha_3).
}
Adding these forms we find
\eq{a-14-11 1formV}{
\delta Q_w + \tilde k_w = -2\delta\left[ \dr\alpha_1 \delta\alpha_1 + \dr\alpha_2 \delta\alpha_2 + (\dr\alpha_3 - z\dr\alpha_2)\delta\alpha_2 \right] + \dr(z\delta\alpha_2 \wedge \delta\alpha_1).
}
Integrating this over the $\phi$ circle the total derivative term does not contribute and we evidently reproduce \eqref{a-14-6 1formV}, from which the phase space action \eqref{a-14-8 1formV} follows.

\section{Relation to Schwazrzian Action for  JT gravity}\label{Relation JT}

From the perspective of the current paper, it is most illuminating to view JT gravity and its Schwarzian action description \cite{Maldacena:2016upp} as a special case of our approach to 3D gravity.  The 3D origin of the JT/Schwarzian is discussed in  \cite{Mertens:2018fds}.

In 3D we have the Euclidean action
\eq{rra}{ S_3& =-\frac{1}{ 16\pi G} \int\! \dr^3x \sqrt{g_3} (R_3+2) -\frac{1}{ 8\pi G} \int_{\p M}\! \dr^2x \sqrt{h_3} (K_3-1)}
The 3D class of metrics we consider are 
\eq{rrb}{ ds_3^2 = \frac{\dr z^2}{ z^2} + \left(\frac{1}{ z^2} +\frac{z^2 }{ 4} L\Lb \right) \dr w \dr \wb -\frac{1}{ 2} L\dr w^2 -\frac{1}{ 2} \Lb \dr\wb^2 }
with $w=\phi+it$, $\wb=\phi-it$, $(L,\Lb)$ are  functions of $(\phi,t)$ that  take the form 
\eq{rrc}{L&= \{F(\phi,t),\phi\} +\frac{\kappa}{2} F'^2 \cr 
\Lb&= \{\Fb(\phi,t),\phi\} +\frac{\overline{\kappa}}{2} \Fb'^2 }
where $(\kappa,\overline{\kappa})$  are constants and $'= \partial_\phi$.  At fixed $t$, the functions $(F,\Fb)$ are elements of diff$(S^1)$.   In general, the metrics \rf{rrb} are off-shell; to be on-shell the functions $(F,\Fb)$ must be, respectively, holomorphic and anti-holomorphic in $w$.   

It is rather tricky to obtain the off-shell action governing \rf{rrb} by direct substitution into \rf{rra}, in part because the coordinates are not globally smooth.   Instead, we apply  phase space methods, viewing $(F,\Fb)$ at fixed time as points on phase space and building an action out of the gravitational symplectic form and Hamiltonian on this phase space.  This procedure \cite{Kraus:2021cwf} leads to the Alekseev-Shatashvili action \cite{Alekseev:1988ce},
\eq{rrd}{
    S_{AS}= -\frac{1}{16 \pi G  } \int \! \dr^2 x \left[  \kappa F' \p_{\bar{w}} F - \left(\frac{1}{F'} \right)^{''} \p_{\bar{w}} F + \bar{\kappa} \bar{F}' \p_{w} \bar{F} - \left(\frac{1}{\bar{F}'} \right)^{''} \p_{w} \bar{F}\right]~. }

We now turn to the 2D story.   One approach is to KK reduce the 3D action by considering metrics 
\eq{rre}{  \dr s_3^2 = \dr s_2^2 +  \Phi^2 \dr t^2~,\quad t \cong t+2\pi }
taking the metric components to be $t$ independent.  Using 
\eq{rrf}{ \int_{M_3}\! \dr^3 x \sqrt{g_3}(R_3-\Lambda)  = \int_{M_2}\! \dr^2x  \sqrt{g_2} \Phi (R_2 -\Lambda)  -2 \int_{\p M_2}\dr x \sqrt{h } n^\mu \p_\mu \Phi  }
where  the metric on the $2d$ boundary is $h$ and $n^\mu$ is the outward pointing unit normal to the boundary, along with 
\eq{rrg}{ \sqrt{g_3} K_3  = \sqrt{h} \Phi K_2 + \sqrt{ h}  n^\nu \p_\mu \Phi }
we find that the 3D action \rf{rrb} reduces to the JT action
\eq{rrh}{ S_3 & = -\frac{1}{ 16 \pi G_2 } \int\! \dr^2x \sqrt{h}\Phi  (R_2+2)  -\frac{1}{ 8\pi G_2 } \int  \! \dr\phi   \sqrt{h} \Phi(K_2-1) }
with $G_2=G/2\pi$. 

Coming back to our 3D picture, to be in accordance with \rf{rre} we should take $F=F(\phi)$ and $\Fb=F(\phi)$, along with $\kappa=\overline{\kappa}$.  We are thus considering 3D metrics (generically off-shell) of the form 
\eq{rri}{ \dr s_3^2   =   \frac{\dr z^2}{ z^2} +    \frac{1}{ z^2}  \left( 1-\frac{Lz^2}{ 2}\right)^2\dr\phi^2 +    \frac{1}{ z^2}  \left(1+\frac{Lz^2}{ 2}\right)^2\dr t^2                }
with $L(\phi) = \{ F(\phi),\phi\} +\frac{\kappa}{ 2} F'^2$.
For this restricted class of off-shell metrics it is straightforward to evaluate the 3D action $S_3$ on the form \rf{rri}.  In particular, we have $R_2 =-2$ so only the boundary term survives.   Cutting off the space at $z=z_c$ and taking $\Phi|_{z=z_c} = z_c^{-1}$, we find the action as $z_c\rt 0$,
\eq{rrj}{ S_3 =  -\frac{1}{8\pi G_2} \int L(\phi) \dr\phi +{\rm constant} }
which is the Schwarzian action (supplemented with the $\kappa$ term).\footnote{The original approach to obtaining this action involved considering cutouts of AdS$_2$ with wiggly boundary.  Here the boundary is at fixed coordinate location $z=z_c$, and wiggles are instead encoded in fluctuations of the metric via the function $L(\phi)$. }     Alternatively, we can apply the reduction to the AS action \rf{rrd}.    After integrating over $t$ and doing some integration by parts,  we find that \rf{rrd} is equal to \rf{rrj}.

\bibliographystyle{bibstyle2017}
\bibliography{collection}

\providecommand{\href}[2]{#2}\begingroup\begin{thebibliography}{10}

\bibitem{Witten:1988hf}
E.~Witten, {\it {Quantum Field Theory and the Jones Polynomial}},
  \href{http://dx.doi.org/10.1007/BF01217730}{{\sf Commun. Math. Phys.} {\sf
  {121} }{\sf (1989) }{\sf 351--399}}.

\bibitem{Elitzur:1989nr}
S.~Elitzur, G.~W. Moore, A.~Schwimmer, and N.~Seiberg, {\it {Remarks on the
  Canonical Quantization of the Chern-Simons-Witten Theory}},
  \href{http://dx.doi.org/10.1016/0550-3213(89)90436-7}{{\sf Nucl. Phys. B}
  {\sf {326} }{\sf (1989) }{\sf 108--134}}.

\bibitem{Wald:1999wa}
R.~M. Wald and A.~Zoupas, {\it {A General definition of 'conserved quantities'
  in general relativity and other theories of gravity}},
  \href{http://dx.doi.org/10.1103/PhysRevD.61.084027}{{\sf Phys. Rev. D} {\sf
  {61} }{\sf (2000) }{\sf 084027}},
  \href{http://arxiv.org/abs/gr-qc/9911095}{{\ttfamily arXiv:gr-qc/9911095}}.

\bibitem{Donnelly:2014fua}
W.~Donnelly and A.~C. Wall, {\it {Entanglement entropy of electromagnetic edge
  modes}},  \href{http://dx.doi.org/10.1103/PhysRevLett.114.111603}{{\sf Phys.
  Rev. Lett.} {\sf {114} }{\sf no.~11, }{\sf (2015) }{\sf 111603}},
  \href{http://arxiv.org/abs/1412.1895}{{\ttfamily arXiv:1412.1895 [hep-th]}}.

\bibitem{Ghosh:2015iwa}
S.~Ghosh, R.~M. Soni, and S.~P. Trivedi, {\it {On The Entanglement Entropy For
  Gauge Theories}},  \href{http://dx.doi.org/10.1007/JHEP09(2015)069}{{\sf
  JHEP} {\sf {09} }{\sf (2015) }{\sf 069}},
  \href{http://arxiv.org/abs/1501.02593}{{\ttfamily arXiv:1501.02593
  [hep-th]}}.

\bibitem{Donnelly:2016auv}
W.~Donnelly and L.~Freidel, {\it {Local subsystems in gauge theory and
  gravity}},  \href{http://dx.doi.org/10.1007/JHEP09(2016)102}{{\sf JHEP} {\sf
  {09} }{\sf (2016) }{\sf 102}},
  \href{http://arxiv.org/abs/1601.04744}{{\ttfamily arXiv:1601.04744
  [hep-th]}}.

\bibitem{Blommaert:2018oue}
A.~Blommaert, T.~G. Mertens, and H.~Verschelde, {\it {Edge dynamics from the
  path integral \textemdash{} Maxwell and Yang-Mills}},
  \href{http://dx.doi.org/10.1007/JHEP11(2018)080}{{\sf JHEP} {\sf {11} }{\sf
  (2018) }{\sf 080}}, \href{http://arxiv.org/abs/1804.07585}{{\ttfamily
  arXiv:1804.07585 [hep-th]}}.

\bibitem{Compere:2015knw}
G.~Comp\`ere, P.~Mao, A.~Seraj, and M.~M. Sheikh-Jabbari, {\it {Symplectic and
  Killing symmetries of AdS$_{3}$ gravity: holographic vs boundary gravitons}},
   \href{http://dx.doi.org/10.1007/JHEP01(2016)080}{{\sf JHEP} {\sf {01} }{\sf
  (2016) }{\sf 080}}, \href{http://arxiv.org/abs/1511.06079}{{\ttfamily
  arXiv:1511.06079 [hep-th]}}.

\bibitem{Seraj:2017rzw}
A.~Seraj and D.~Van~den Bleeken, {\it {Strolling along gauge theory vacua}},
  \href{http://dx.doi.org/10.1007/JHEP08(2017)127}{{\sf JHEP} {\sf {08} }{\sf
  (2017) }{\sf 127}}, \href{http://arxiv.org/abs/1707.00006}{{\ttfamily
  arXiv:1707.00006 [hep-th]}}.

\bibitem{Kutluk:2019ghr}
E.~c. Kutluk, A.~Seraj, and D.~Van Den~Bleeken, {\it {Strolling along
  gravitational vacua}},  \href{http://dx.doi.org/10.1007/JHEP01(2020)184}{{\sf
  JHEP} {\sf {01} }{\sf (2020) }{\sf 184}},
  \href{http://arxiv.org/abs/1904.12869}{{\ttfamily arXiv:1904.12869
  [hep-th]}}.

\bibitem{Adami:2020ugu}
H.~Adami, M.~M. Sheikh-Jabbari, V.~Taghiloo, H.~Yavartanoo, and C.~Zwikel, {\it
  {Symmetries at null boundaries: two and three dimensional gravity cases}},
  \href{http://dx.doi.org/10.1007/JHEP10(2020)107}{{\sf JHEP} {\sf {10} }{\sf
  (2020) }{\sf 107}}, \href{http://arxiv.org/abs/2007.12759}{{\ttfamily
  arXiv:2007.12759 [hep-th]}}.

\bibitem{Adami:2021sko}
H.~Adami, M.~M. Sheikh-Jabbari, V.~Taghiloo, H.~Yavartanoo, and C.~Zwikel, {\it
  {Chiral Massive News: Null Boundary Symmetries in Topologically Massive
  Gravity}},  \href{http://dx.doi.org/10.1007/JHEP05(2021)261}{{\sf JHEP} {\sf
  {05} }{\sf (2021) }{\sf 261}},
  \href{http://arxiv.org/abs/2104.03992}{{\ttfamily arXiv:2104.03992
  [hep-th]}}.

\bibitem{Crnkovic:1986ex}
C.~Crnkovic and E.~Witten, {\it {Covariant description of canonical formalism
  in geometrical theories}},  {\sf Three hundred years of gravitation} {\sf
  (1986) }{\sf 676--684}.

\bibitem{Lee:1990nz}
J.~Lee and R.~M. Wald, {\it {Local symmetries and constraints}},
  \href{http://dx.doi.org/10.1063/1.528801}{{\sf J. Math. Phys.} {\sf {31}
  }{\sf (1990) }{\sf 725--743}}.

\bibitem{Achucarro:1986uwr}
A.~Achucarro and P.~K. Townsend, {\it {A Chern-Simons Action for
  Three-Dimensional anti-De Sitter Supergravity Theories}},
  \href{http://dx.doi.org/10.1016/0370-2693(86)90140-1}{{\sf Phys. Lett. B}
  {\sf {180} }{\sf (1986) }{\sf 89}}.

\bibitem{Witten:1988hc}
E.~Witten, {\it {(2+1)-Dimensional Gravity as an Exactly Soluble System}},
  \href{http://dx.doi.org/10.1016/0550-3213(88)90143-5}{{\sf Nucl. Phys. B}
  {\sf {311} }{\sf (1988) }{\sf 46}}.

\bibitem{Coussaert:1995zp}
O.~Coussaert, M.~Henneaux, and P.~van Driel, {\it {The Asymptotic dynamics of
  three-dimensional Einstein gravity with a negative cosmological constant}},
  \href{http://dx.doi.org/10.1088/0264-9381/12/12/012}{{\sf Class. Quant.
  Grav.} {\sf {12} }{\sf (1995) }{\sf 2961--2966}},
  \href{http://arxiv.org/abs/gr-qc/9506019}{{\ttfamily arXiv:gr-qc/9506019}}.

\bibitem{Cotler:2018zff}
J.~Cotler and K.~Jensen, {\it {A theory of reparameterizations for AdS$_3$
  gravity}},  \href{http://dx.doi.org/10.1007/JHEP02(2019)079}{{\sf JHEP} {\sf
  {02} }{\sf (2019) }{\sf 079}},
  \href{http://arxiv.org/abs/1808.03263}{{\ttfamily arXiv:1808.03263
  [hep-th]}}.

\bibitem{Barnich:2017jgw}
G.~Barnich, H.~A. Gonzalez, and P.~Salgado-Rebolledo, {\it {Geometric actions
  for three-dimensional gravity}},
  \href{http://dx.doi.org/10.1088/1361-6382/aa9806}{{\sf Class. Quant. Grav.}
  {\sf {35} }{\sf no.~1, }{\sf (2018) }{\sf 014003}},
  \href{http://arxiv.org/abs/1707.08887}{{\ttfamily arXiv:1707.08887
  [hep-th]}}.

\bibitem{Kraus:2021cwf}
P.~Kraus, R.~Monten, and R.~M. Myers, {\it {3D Gravity in a Box}},
  \href{http://dx.doi.org/10.21468/SciPostPhys.11.3.070}{{\sf SciPost Phys.}
  {\sf {11} }{\sf (2021) }{\sf 070}},
  \href{http://arxiv.org/abs/2103.13398}{{\ttfamily arXiv:2103.13398
  [hep-th]}}.

\bibitem{Ebert:2022cle}
S.~Ebert, E.~Hijano, P.~Kraus, R.~Monten, and R.~M. Myers, {\it {Field Theory
  of Interacting Boundary Gravitons}},
  \href{http://dx.doi.org/10.21468/SciPostPhys.13.2.038}{{\sf SciPost Phys.}
  {\sf {13} }{\sf no.~2, }{\sf (2022) }{\sf 038}},
  \href{http://arxiv.org/abs/2201.01780}{{\ttfamily arXiv:2201.01780
  [hep-th]}}.

\bibitem{Anninos:2008fx}
D.~Anninos, W.~Li, M.~Padi, W.~Song, and A.~Strominger, {\it {Warped AdS(3)
  Black Holes}},  \href{http://dx.doi.org/10.1088/1126-6708/2009/03/130}{{\sf
  JHEP} {\sf {03} }{\sf (2009) }{\sf 130}},
  \href{http://arxiv.org/abs/0807.3040}{{\ttfamily arXiv:0807.3040 [hep-th]}}.

\bibitem{Deser:1981wh}
S.~Deser, R.~Jackiw, and S.~Templeton, {\it {Topologically Massive Gauge
  Theories}},  \href{http://dx.doi.org/10.1016/0003-4916(82)90164-6}{{\sf
  Annals Phys.} {\sf {140} }{\sf (1982) }{\sf 372--411}}. [Erratum: Annals
  Phys. 185, 406 (1988)].

\bibitem{Kraus:2022mnu}
P.~Kraus, R.~Monten, and K.~Roumpedakis, {\it {Refining the cutoff 3d gravity/$
  T\overline{T} $ correspondence}},
  \href{http://dx.doi.org/10.1007/JHEP10(2022)094}{{\sf JHEP} {\sf {10} }{\sf
  (2022) }{\sf 094}}, \href{http://arxiv.org/abs/2206.00674}{{\ttfamily
  arXiv:2206.00674 [hep-th]}}.

\bibitem{Nutku:1993eb}
Y.~Nutku, {\it {Exact solutions of topologically massive gravity with a
  cosmological constant}},
  \href{http://dx.doi.org/10.1088/0264-9381/10/12/022}{{\sf Class. Quant.
  Grav.} {\sf {10} }{\sf (1993) }{\sf 2657--2661}}.

\bibitem{Gurses:1994bjn}
M.~G\"urses, {\it {Perfect Fluid Sources in 2+1 Dimensions}},
  \href{http://dx.doi.org/10.1088/0264-9381/11/10/017}{{\sf Class. Quant.
  Grav.} {\sf {11} }{\sf no.~10, }{\sf (1994) }{\sf 2585}}.

\bibitem{Hofman:2014loa}
D.~M. Hofman and B.~Rollier, {\it {Warped Conformal Field Theory as Lower Spin
  Gravity}},  \href{http://dx.doi.org/10.1016/j.nuclphysb.2015.05.011}{{\sf
  Nucl. Phys. B} {\sf {897} }{\sf (2015) }{\sf 1--38}},
  \href{http://arxiv.org/abs/1411.0672}{{\ttfamily arXiv:1411.0672 [hep-th]}}.

\bibitem{Maloney:2007ud}
A.~Maloney and E.~Witten, {\it {Quantum Gravity Partition Functions in Three
  Dimensions}},  \href{http://dx.doi.org/10.1007/JHEP02(2010)029}{{\sf JHEP}
  {\sf {02} }{\sf (2010) }{\sf 029}},
  \href{http://arxiv.org/abs/0712.0155}{{\ttfamily arXiv:0712.0155 [hep-th]}}.

\bibitem{Compere:2018aar}
G.~Comp\`ere and A.~Fiorucci, {\it {Advanced Lectures on General Relativity}},
  \href{http://arxiv.org/abs/1801.07064}{{\ttfamily arXiv:1801.07064
  [hep-th]}}.

\bibitem{Harlow:2019yfa}
D.~Harlow and J.-Q. Wu, {\it {Covariant phase space with boundaries}},
  \href{http://dx.doi.org/10.1007/JHEP10(2020)146}{{\sf JHEP} {\sf {10} }{\sf
  (2020) }{\sf 146}}, \href{http://arxiv.org/abs/1906.08616}{{\ttfamily
  arXiv:1906.08616 [hep-th]}}.

\bibitem{Wald:1993nt}
R.~M. Wald, {\it {Black hole entropy is the Noether charge}},
  \href{http://dx.doi.org/10.1103/PhysRevD.48.R3427}{{\sf Phys. Rev. D} {\sf
  {48} }{\sf no.~8, }{\sf (1993) }{\sf R3427--R3431}},
  \href{http://arxiv.org/abs/gr-qc/9307038}{{\ttfamily arXiv:gr-qc/9307038}}.

\bibitem{Tachikawa:2006sz}
Y.~Tachikawa, {\it {Black hole entropy in the presence of Chern-Simons terms}},
   \href{http://dx.doi.org/10.1088/0264-9381/24/3/014}{{\sf Class. Quant.
  Grav.} {\sf {24} }{\sf (2007) }{\sf 737--744}},
  \href{http://arxiv.org/abs/hep-th/0611141}{{\ttfamily arXiv:hep-th/0611141}}.

\bibitem{Coleman:1969sm}
S.~R. Coleman, J.~Wess, and B.~Zumino, {\it {Structure of phenomenological
  Lagrangians. 1.}},  \href{http://dx.doi.org/10.1103/PhysRev.177.2239}{{\sf
  Phys. Rev.} {\sf {177} }{\sf (1969) }{\sf 2239--2247}}.

\bibitem{Callan:1969sn}
C.~G. Callan, Jr., S.~R. Coleman, J.~Wess, and B.~Zumino, {\it {Structure of
  phenomenological Lagrangians. 2.}},
  \href{http://dx.doi.org/10.1103/PhysRev.177.2247}{{\sf Phys. Rev.} {\sf {177}
  }{\sf (1969) }{\sf 2247--2250}}.

\bibitem{Compere:2013bya}
G.~Comp\`ere, W.~Song, and A.~Strominger, {\it {New Boundary Conditions for
  AdS3}},  \href{http://dx.doi.org/10.1007/JHEP05(2013)152}{{\sf JHEP} {\sf
  {05} }{\sf (2013) }{\sf 152}},
  \href{http://arxiv.org/abs/1303.2662}{{\ttfamily arXiv:1303.2662 [hep-th]}}.

\bibitem{Alekseev:1988ce}
A.~Alekseev and S.~L. Shatashvili, {\it {Path Integral Quantization of the
  Coadjoint Orbits of the Virasoro Group and 2D Gravity}},
  \href{http://dx.doi.org/10.1016/0550-3213(89)90130-2}{{\sf Nucl. Phys. B}
  {\sf {323} }{\sf (1989) }{\sf 719--733}}.

\bibitem{Banados:1998gg}
M.~Ba\~nados, {\it {Three-dimensional quantum geometry and black holes}},
  \href{http://dx.doi.org/10.1063/1.59661}{{\sf AIP Conf. Proc.} {\sf {484}
  }{\sf no.~1, }{\sf (1999) }{\sf 147--169}},
  \href{http://arxiv.org/abs/hep-th/9901148}{{\ttfamily arXiv:hep-th/9901148}}.

\bibitem{Iyer:1994ys}
V.~Iyer and R.~M. Wald, {\it {Some properties of Noether charge and a proposal
  for dynamical black hole entropy}},
  \href{http://dx.doi.org/10.1103/PhysRevD.50.846}{{\sf Phys. Rev. D} {\sf {50}
  }{\sf (1994) }{\sf 846--864}},
  \href{http://arxiv.org/abs/gr-qc/9403028}{{\ttfamily arXiv:gr-qc/9403028}}.

\bibitem{Nazaroglu:2011zi}
C.~Nazaroglu, Y.~Nutku, and B.~Tekin, {\it {Covariant Symplectic Structure and
  Conserved Charges of Topologically Massive Gravity}},
  \href{http://dx.doi.org/10.1103/PhysRevD.83.124039}{{\sf Phys. Rev. D} {\sf
  {83} }{\sf (2011) }{\sf 124039}},
  \href{http://arxiv.org/abs/1104.3404}{{\ttfamily arXiv:1104.3404 [hep-th]}}.

\bibitem{Bouchareb:2007yx}
A.~Bouchareb and G.~Clement, {\it {Black hole mass and angular momentum in
  topologically massive gravity}},
  \href{http://dx.doi.org/10.1088/0264-9381/24/22/018}{{\sf Class. Quant.
  Grav.} {\sf {24} }{\sf (2007) }{\sf 5581--5594}},
  \href{http://arxiv.org/abs/0706.0263}{{\ttfamily arXiv:0706.0263 [gr-qc]}}.

\bibitem{Compere:2008cv}
G.~Compere and S.~Detournay, {\it {Semi-classical central charge in
  topologically massive gravity}},
  \href{http://dx.doi.org/10.1088/0264-9381/26/1/012001}{{\sf Class. Quant.
  Grav.} {\sf {26} }{\sf (2009) }{\sf 012001}},
  \href{http://arxiv.org/abs/0808.1911}{{\ttfamily arXiv:0808.1911 [hep-th]}}.
  [Erratum: Class.Quant.Grav. 26, 139801 (2009)].

\bibitem{Compere:2009zj}
G.~Compere and S.~Detournay, {\it {Boundary conditions for spacelike and
  timelike warped $AdS_{3}$ spaces in topologically massive gravity}},
  \href{http://dx.doi.org/10.1088/1126-6708/2009/08/092}{{\sf JHEP} {\sf {08}
  }{\sf (2009) }{\sf 092}}, \href{http://arxiv.org/abs/0906.1243}{{\ttfamily
  arXiv:0906.1243 [hep-th]}}.

\bibitem{Blagojevic:2009ek}
M.~Blagojevic and B.~Cvetkovic, {\it {Asymptotic structure of topologically
  massive gravity in spacelike stretched AdS sector}},
  \href{http://dx.doi.org/10.1088/1126-6708/2009/09/006}{{\sf JHEP} {\sf {09}
  }{\sf (2009) }{\sf 006}}, \href{http://arxiv.org/abs/0907.0950}{{\ttfamily
  arXiv:0907.0950 [gr-qc]}}.

\bibitem{Azeyanagi:2018har}
T.~Azeyanagi, S.~Detournay, and M.~Riegler, {\it {Warped Black Holes in
  Lower-Spin Gravity}},
  \href{http://dx.doi.org/10.1103/PhysRevD.99.026013}{{\sf Phys. Rev. D} {\sf
  {99} }{\sf no.~2, }{\sf (2019) }{\sf 026013}},
  \href{http://arxiv.org/abs/1801.07263}{{\ttfamily arXiv:1801.07263
  [hep-th]}}.

\bibitem{Detournay:2012pc}
S.~Detournay, T.~Hartman, and D.~M. Hofman, {\it {Warped Conformal Field
  Theory}},  \href{http://dx.doi.org/10.1103/PhysRevD.86.124018}{{\sf Phys.
  Rev. D} {\sf {86} }{\sf (2012) }{\sf 124018}},
  \href{http://arxiv.org/abs/1210.0539}{{\ttfamily arXiv:1210.0539 [hep-th]}}.

\bibitem{DHoker:2009mmn}
E.~D'Hoker and P.~Kraus, {\it {Magnetic Brane Solutions in AdS}},
  \href{http://dx.doi.org/10.1088/1126-6708/2009/10/088}{{\sf JHEP} {\sf {10}
  }{\sf (2009) }{\sf 088}}, \href{http://arxiv.org/abs/0908.3875}{{\ttfamily
  arXiv:0908.3875 [hep-th]}}.

\bibitem{DHoker:2010xwl}
E.~D'Hoker, P.~Kraus, and A.~Shah, {\it {RG Flow of Magnetic Brane
  Correlators}},  \href{http://dx.doi.org/10.1007/JHEP04(2011)039}{{\sf JHEP}
  {\sf {04} }{\sf (2011) }{\sf 039}},
  \href{http://arxiv.org/abs/1012.5072}{{\ttfamily arXiv:1012.5072 [hep-th]}}.

\bibitem{Adawi:1998ta}
T.~Adawi, M.~Cederwall, U.~Gran, B.~E.~W. Nilsson, and B.~Razaznejad, {\it
  {Goldstone tensor modes}},
  \href{http://dx.doi.org/10.1088/1126-6708/1999/02/001}{{\sf JHEP} {\sf {02}
  }{\sf (1999) }{\sf 001}},
  \href{http://arxiv.org/abs/hep-th/9811145}{{\ttfamily arXiv:hep-th/9811145}}.

\bibitem{Cederwall:1998tr}
M.~Cederwall, U.~Gran, M.~Holm, and B.~E.~W. Nilsson, {\it {Finite tensor
  deformations of supergravity solitons}},
  \href{http://dx.doi.org/10.1088/1126-6708/1999/02/003}{{\sf JHEP} {\sf {02}
  }{\sf (1999) }{\sf 003}},
  \href{http://arxiv.org/abs/hep-th/9812144}{{\ttfamily arXiv:hep-th/9812144}}.

\bibitem{Wald:1990}
R.~M. Wald, {\it On identically closed forms locally constructed from a field},
   \href{http://dx.doi.org/10.1063/1.528839}{{\sf Journal of Mathematical
  Physics} {\sf {31} }{\sf no.~10, }{\sf (1990) }{\sf 2378--2384}},
  \href{http://arxiv.org/abs/https://doi.org/10.1063/1.528839}{{\ttfamily
  https://doi.org/10.1063/1.528839}}. \url{https://doi.org/10.1063/1.528839}.

\bibitem{Barnich:2000zw}
G.~Barnich, F.~Brandt, and M.~Henneaux, {\it {Local BRST cohomology in gauge
  theories}},  \href{http://dx.doi.org/10.1016/S0370-1573(00)00049-1}{{\sf
  Phys. Rept.} {\sf {338} }{\sf (2000) }{\sf 439--569}},
  \href{http://arxiv.org/abs/hep-th/0002245}{{\ttfamily arXiv:hep-th/0002245}}.

\bibitem{Barnich:2001jy}
G.~Barnich and F.~Brandt, {\it {Covariant theory of asymptotic symmetries,
  conservation laws and central charges}},
  \href{http://dx.doi.org/10.1016/S0550-3213(02)00251-1}{{\sf Nucl. Phys. B}
  {\sf {633} }{\sf (2002) }{\sf 3--82}},
  \href{http://arxiv.org/abs/hep-th/0111246}{{\ttfamily arXiv:hep-th/0111246}}.

\bibitem{Olver:1986}
P.~J. Olver, \href{http://dx.doi.org/10.1007/978-1-4684-0274-2}{{\it
  {Applications of Lie Groups to Differential Equations}}, }.
\newblock Graduate Texts in Mathematics. Springer New York, NY, 1986.

\bibitem{Anderson:1989}
I.~M. Anderson, {\it {The Variational Bicomplex}}, .
\newblock Utah State University, 1989.

\bibitem{Compere:2007az}
G.~Compere, {\it {Symmetries and conservation laws in Lagrangian gauge theories
  with applications to the mechanics of black holes and to gravity in three
  dimensions}}, .
\newblock PhD thesis, Brussels U., 2007.
\newblock \href{http://arxiv.org/abs/0708.3153}{{\ttfamily arXiv:0708.3153
  [hep-th]}}.

\bibitem{Maldacena:2016upp}
J.~Maldacena, D.~Stanford, and Z.~Yang, {\it {Conformal symmetry and its
  breaking in two dimensional Nearly Anti-de-Sitter space}},
  \href{http://dx.doi.org/10.1093/ptep/ptw124}{{\sf PTEP} {\sf {2016} }{\sf
  no.~12, }{\sf (2016) }{\sf 12C104}},
  \href{http://arxiv.org/abs/1606.01857}{{\ttfamily arXiv:1606.01857
  [hep-th]}}.

\bibitem{Mertens:2018fds}
T.~G. Mertens, {\it {The Schwarzian theory \textemdash{} origins}},
  \href{http://dx.doi.org/10.1007/JHEP05(2018)036}{{\sf JHEP} {\sf {05} }{\sf
  (2018) }{\sf 036}}, \href{http://arxiv.org/abs/1801.09605}{{\ttfamily
  arXiv:1801.09605 [hep-th]}}.

\end{thebibliography}\endgroup

\end{document}